# Spin Resonance Amplitude and Frequency of a Single Atom on a Surface in a Vector Magnetic Field

Jinkyung Kim,[1,2,†] Won-jun Jang,[4,†] Thi Hong Bui,[1,2] Deung-jang Choi,[5,6,7] Christoph Wolf,[1,3] Fernando Delgado,[8] Denis Krylov,[1,3] Soonhyeong Lee,[1,3] Sangwon Yoon,[1,3] Christopher P. Lutz,[9] Andreas J. Heinrich,[1,2,*] and Yujeong Bae[1,2,*]

[1]Center for Quantum Nanoscience (QNS), Institute for Basic Science (IBS), Seoul 03760, South Korea

[2]Department of Physics, Ewha Womans University, Seoul 03760, South Korea

[3]Ewha Womans University, Seoul 03760, Republic of Korea

[4]Nano Electronics Lab, Samsung Advanced Institute of Technology, Suwon 13595, South Korea

[5]Centro de Física de Materiales CFM/MPC (CSIC-UPV/EHU), 20018 San Sebastián, Spain

[6]Donostia International Physics Center (DIPC), 20018 Donostia-San Sebastian, Spain

[7]Ikerbasque, Basque Foundation for Science, 48013 Bilbao, Spain

[8]Instituto de estudios avanzados IUDEA, Departamento de Fisica, Universidad de La Laguna, 38203, Tenerife, Spain

[9]IBM Almaden Research Center, San Jose, CA 95120, USA

[†]These authors contributed equally to this work.

*Corresponding authors: A.J.H. (heinrich.andreas@qns.science), Y.B. (bae.yujeong@qns.science)

**We used electron spin resonance (ESR) combined with scanning tunneling microscopy (STM) to measure hydrogenated Ti (spin-1/2) atoms at low-symmetry binding sites on MgO in vector magnetic fields. We found strongly anisotropic g-values in all three spatial directions. Interestingly, the amplitude and lineshape of the ESR signals are also strongly dependent on the angle of the field. We conclude that the Ti spin is aligned along the magnetic field, while the tip spin follows its strong magnetic anisotropy. Our results show the interplay between the tip and surface spins in determining the ESR signals and highlight the precision of ESR-STM to identify the single atom's spin states.**

Electron spin resonance (ESR) offers high energy resolution for the measurement of spins, which is not limited by temperature. However, it usually requires a very large number of identical spins to achieve sufficient signal [1] and, thus, provides averaged information on an ensemble of spins. Scanning tunneling microscopy (STM), on the other hand, offers access to individual spins and to the surrounding environment at the atomic scale, albeit with an energy resolution that is limited by the temperature of tip and sample [2]. Recently, it was shown that these advantages could be combined in ESR-STM, which achieves an energy resolution of around 10 nano-electronvolt with atomic-scale spatial resolution on individual spins [3,4].

The most common type of spin centers in ensemble ESR has the spin of $S = 1/2$ and measuring the dependence of the Zeeman energy on both the magnitude and the direction of magnetic fields lies at the core of ESR. A well-studied spin-1/2 system in ESR-STM is a Ti atom on a thin MgO film supported on Ag(100) [5-7], where Ti atoms are found at two different binding sites and are presumably hydrogenated [5,7,8]. Among the intensive reports of ESR-STM on Ti atoms, there was a discrepancy in the reported g-values [6,7,9,10]. Unlike ensemble ESR, ESR-STM rarely has the capability to change the direction of magnetic fields, which makes it difficult to unravel the origin of anisotropic g-values. It was recently found that such Ti at the oxygen binding site has a very highly anisotropic g-factor with the in-plane component being 2.7 times larger than the out-of-plane one [10]. Due to the 4-fold symmetry of the O binding site, the two in-plane components of the g-factor have to be identical. Initial measurements also hinted at an anisotropic g-factor on the lower symmetry (2-fold) bridge binding site on the same substrate [7,9].

Three different transition metal atoms have been studied using ESR-STM [3,5,11] and several mechanisms have been proposed to explain the coherent driving and the detection of spins in ESR-STM. Proposed driving mechanisms include the physical motion of the ESR-active spin in the presence of an inhomogeneous magnetic field created by the tip, also referred to as piezoelectric coupling (PEC) [6,12], as well as a modulation of the tunneling barrier by the applied AC voltage [13]. Recently, strongly persuasive results of ESR measurements based on the PEC model have been reported [6,14].

Exploring ESR as a function of the direction of an applied magnetic field can shed light on an anisotropic g-factor [10] as well as the ESR driving mechanism. Here, we utilize a two-fold symmetric binding site of

Ti on MgO [6] and vary the direction of the magnetic field in small increments in a two-dimensional plane (*xz*-plane) [Fig. 1(a)]. These measurements were taken with a home-built ESR-STM system operating at 1K (Supplemental Sec. 1).

We deposited Ti and Fe atoms on a two atomic layer film of MgO grown on an Ag(100) substrate [Fig. 1(b)]. Ti and Fe atoms can be readily distinguished by scanning tunneling spectroscopy with inelastic electron tunneling spectroscopy steps [2]. We found two different types of Ti atoms at bridge binding sites as shown in STM topographic images [Fig. 1(b)]. The difference originates from the relative angle of the MgO lattice with the Ti spin [7,9]. The spin direction of Ti follows the external magnetic field since Ti has no higher-order anisotropy terms in its spin Hamiltonian [5]. We label Ti at the two different bridge binding sites as vertical $Ti_v$ and horizontal $Ti_h$. For $Ti_v$, the nearest neighboring oxygen atoms are almost aligned with the in-plane component of the external magnetic field, while they are almost perpendicular in the case of $Ti_h$ [Fig. 1(b), inset].

An example of a typical ESR spectrum for Ti at the bridge binding site is shown in Fig. 1(c), measured by sweeping the frequency of a radio frequency voltage ($V_{RF}$), $f_{RF}$, at a constant external magnetic field $B_{ext}$. The ESR data were fitted to Eq. (1),

$$I = I_0 + I_1 \cdot \frac{1+\alpha\delta}{1+\delta^2}, \qquad (1)$$

where $\delta$ is the normalized frequency ($\delta = \frac{f-f_{res}}{\Gamma/2}$) for the resonance frequency $f_{res}$ and the linewidth $\Gamma$, $I_1$ is the amplitude of the current change of the ESR signal, $I_0$ is the background current, and $\alpha$ is the asymmetry factor [6]. When $\alpha$ is zero, Eq. (1) corresponds to the Lorentzian function. From the example fit in Fig. 1(c), we obtained a resonance frequency $f_{res} = 18.361 \pm 0.001$ GHz and the amplitude of the spectrum $I_1 = 835 \pm 11$ fA with $B_{ext} = 0.7$ T out of a DC current of $I_{set} = 20$ pA.

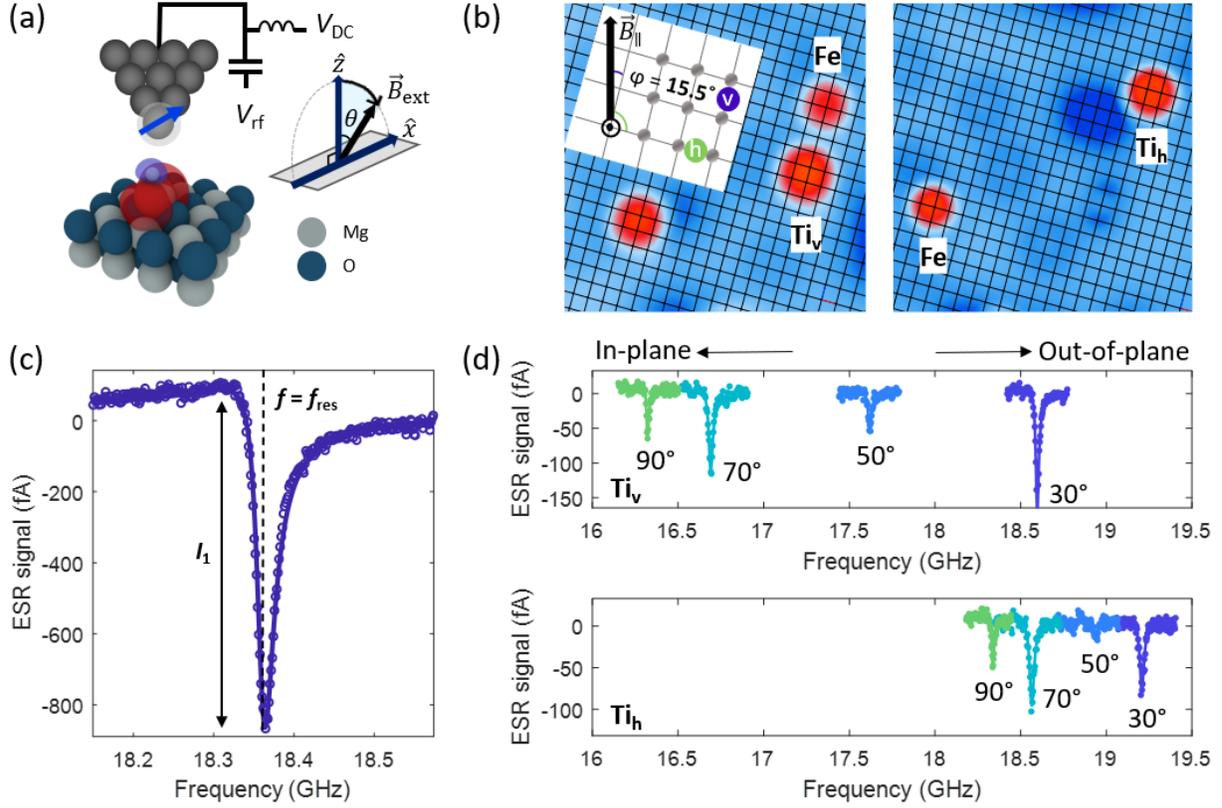

FIG. 1. Hydrogenated-Ti atom on a bridge binding site (Ti$_B$) studied by ESR-STM. (a) Schematic of the ESR-STM set up combined with the calculated spin density of Ti$_B$ on MgO by DFT. The external magnetic field rotates in a two-dimensional plane ($xz$-plane) with the angle $\theta$. Here, $\theta$ is zero when the external magnetic field is in the out-of-plane direction of the sample ($z$-direction). (b) STM images of Ti$_B$ atoms that have different binding configurations on MgO ($V_{DC}$ = 100 mV, $I_{set}$ = 20 pA (left), 10 pA (right), and $T$ = 1.12 K). The intercepts of grid lines represent the positions of oxygen atoms of the MgO lattice. The grid lines are adjusted using Ti and Fe atoms on the top of oxygen as markers. Inset: The angle between the in-plane direction of the external magnetic field ($B_\parallel$) and a line connecting two neighboring oxygen atoms is $\varphi$. When $\varphi$ = 15.5°, Ti$_B$ atoms are labelled as Ti$_v$. In the other case ($\varphi$ = 105.5°), they are labelled as Ti$_h$. Note that $\varphi$ = 105.5° (Ti$_h$) can be re-marked as $\varphi$ = 74.5° due to the two-fold symmetry of Ti$_B$. (c) ESR spectrum of Ti$_v$ at $B_{ext}$ = 0.7 T and $\theta$ = 90° ($V_{DC}$ = 40 mV, $I_{set}$ = 20 pA, $V_{RF}$ = 10 mV and $T$ = 1.12 K). The solid line is the best fit to Eq. (1). (d) ESR spectra of Ti$_v$ (upper) and Ti$_h$ (lower) at different angles $\theta$ of a vector magnetic field (same field magnitude) ($V_{DC}$ = 40 mV, $V_{RF}$ = 10 mV, $I_{set}$ = 20 pA, $B_{ext}$ = 0.7 T, and $T$ = 1.12 K).

In the following, we compare ESR spectra measured on Ti at two different bridge binding sites. Note that we define $\theta$ as the angle between the surface normal and the magnetic field direction. We vary the angle $\theta$ as well as the magnitude of the external magnetic field ($B_{ext}$) in a two-dimensional plane. Interestingly, we found a pronounced difference in the ESR spectra depending on the angle $\theta$ when measuring on $Ti_v$ and $Ti_h$ [Fig. 1(d)]. The resonance frequencies of both $Ti_v$ and $Ti_h$ increased and became more similar as we applied the magnetic field closer to the out-of-plane direction. The observed shift of the ESR frequency of $Ti_v$ with the angle $\theta$ of the magnetic field is almost three times larger than $Ti_h$.

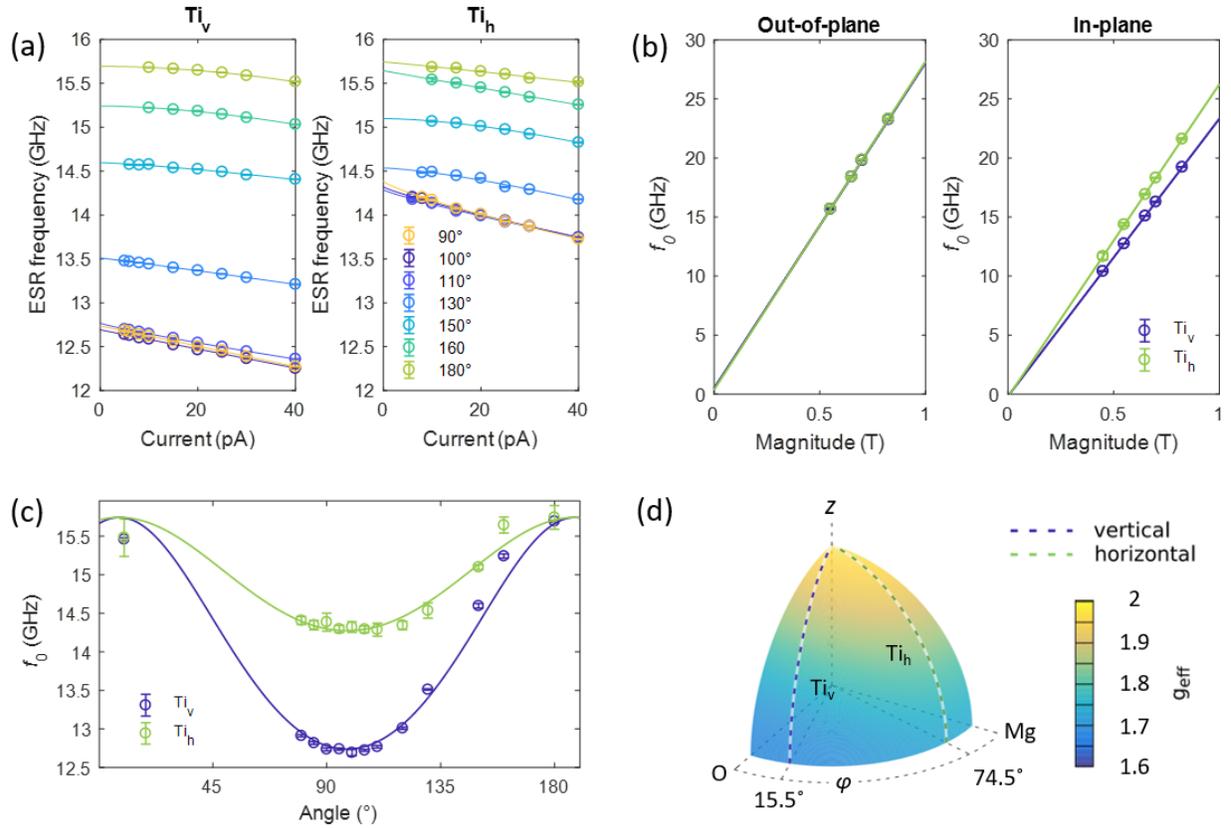

FIG. 2. Anisotropic g-factor of $Ti_B$. (a) Current dependence of the resonance frequency of $Ti_B$. Solid lines are fits considering the exchange interaction between STM tip and $Ti_B$. We give error bars with 95% confidence interval ($V_{DC}$ = 40 mV, $V_{RF}$ = 10 mV, $T$ = 1.12 K, $B_{ext}$ = 0.55 T). (b) ESR frequency extrapolated to zero current ($f_0$) at different magnitudes of $B_{ext}$. We obtained the g-factors of $Ti_v$ and $Ti_h$, $g_{v\parallel}$ = 1.679 ± 0.003 and $g_{h\parallel}$ = 1.901 ± 0.020 at the in-plane field direction (left) and $g_{v\perp}$ = 1.975 ± 0.011 and $g_{h\perp}$ = 1.991 ± 0.019 at the out-of-plane field direction

(right) ($V_{DC}$ = 40 mV, $V_{RF}$ = 10 mV, $T$ = 1.12 K). We note that the fit has a small offset near zero, which might come from a remaining tip effect of the extrapolated $f_0$. (c) $f_0$ as a function of the angle $\theta$. Each curve is fit to an ellipse in the $xz$-plane, tracing along the shape of the two-dimensional anisotropic $g$-factor. Error bars are from the fit in (a). (d) Effective g-factor plotted as a function of the direction of $\boldsymbol{B}_{ext}$ in real space as obtained from the multiplet calculations. The symmetry directions for $Ti_v$ and $Ti_h$ are indicated as slices along the dotted lines, considering the symmetry of the MgO lattice.

In order to precisely measure the dependence on the angle of the magnetic field, it is necessary to measure the ESR spectra at different tip-Ti distances in order to cancel the effective magnetic field created by the spin-polarized tip [15]. At a given $\boldsymbol{B}_{ext}$, the ESR resonance frequency shifts with tip-atom separation ($d$) [Fig. 2(a)]. This shift is governed by two distinct terms of the tip field [5,16], an exchange field, $B_{exc} \propto \exp(-d/d_{ex})$, and a dipolar field, $B_{dip} \propto 1/d^3$. We find that the exchange field dominates for the tip used here as in several previous studies [6,14]. Extrapolating the resonance frequency $f_0$ at zero tunnel current allows us to determine the Zeeman energy corresponding only to the external magnetic field $\boldsymbol{B}_{ext}$.

The extrapolated resonance frequency $f_0$ is plotted as a function of the field angle $\theta$ in Fig. 2(c), where we keep the magnitude of $\boldsymbol{B}_{ext}$ = 0.55 T constant. Near the out-of-plane direction ($\theta$ = 0°), $Ti_v$ and $Ti_h$ show the same ESR frequency. They start shifting differently when we change the angle $\theta$ and, eventually, differ by about 1.54 GHz at the in-plane magnetic field. Our modeling suggests that $Ti_v$ and $Ti_h$ sense different components of the in-plane magnetic field due to different orbital distributions with respect to the external magnetic field. The dependence of $f_0$ on the angle $\theta$ can be fitted to an ellipse, corresponding to the cross-section of the three-dimensional shape of the anisotropic g-factor.

To accurately determine the g-factor along the three principal axes, we measured $f_0$ as a function of magnetic field magnitude $B_{ext}$ for both $Ti_v$ and $Ti_h$ at angles $\theta$ = 0° and 90° of the magnetic field [Fig. 2(b)]. From these measurements, the magnetic moments of the two Ti atoms were obtained from the linear fit of $f_0$ as a function of $B_{ext}$, which yields in-plane magnetic moments $\mu_{v\parallel} = (0.840 \pm 0.001)\mu_B$ and $\mu_{h\parallel} = (0.950 \pm 0.010)\mu_B$ for $Ti_v$ and $Ti_h$, respectively. The out-of-plane magnetic moments for $Ti_v$ and $Ti_h$ are

$\mu_{v\perp} = (0.987 \pm 0.006)\mu_B$ and $\mu_{h\perp} = (0.996 \pm 0.010)\mu_B$. Given that in our sample the O-O bonding in MgO lattices is misaligned by 15.5° about the in-plane component of the external magnetic field [Fig. 1(b)], we can determine the g-factors along the three principal axes: $g_{O-O} = 1.660 \pm 0.002$ (along the nearest neighbor O atoms), $g_{Mg-Mg} = 1.916 \pm 0.018$ and $g_z = 1.984 \pm 0.007$. We confirmed that the result of an anisotropic g-factor is reliable with different tips as well as on other Ti atoms (see Supplemental Sec. 6).

We calculated the effective g-factor along arbitrary directions of the external magnetic field $\bm{B}_{ext}$ [Fig. 2(d)] using a combination of density functional theory (DFT) and magnetic multiplet simulations [17]. Details about the calculation as well as the electronic structure can be found in Supplemental Sec. 5. Following previous studies [5,10], we modeled the system by relaxing a Ti atom on a bridge position between two neighboring oxygen atoms of 2ML MgO and then added a single hydrogen on top of the Ti atom. The ionic coordinates and partial charges of the crystal field were then extracted after an ionic relaxation. In the multiplet calculations of the effective g-factor, we found that a contribution of the orbital moment is less than $0.1\mu_B$ along the $\varphi = 0°$ direction, which couples antiferromagnetically with the spin moment. The presence of orbital momentum strongly affects the spin momentum through the spin-orbit coupling, which causes the anisotropy of the g-factor. The multiplet calculations reproduce the experimental findings to a high degree (see TABLE I). The only significant deviation occurs for $g_{Mg-Mg}$, which could be attributed to effects stemming from covalency. There is a back-donation effect only from Mg atom, and this attribution is not well captured in our point-charge crystal field approach.

TABLE I. Values for the g-factor from experiment and multiplet calculations

| g-factor element | Experiment | Multiplet |
| --- | --- | --- |
| $g_{O-O}$ | $1.660 \pm 0.002$ | 1.68 |
| $g_{Mg-Mg}$ | $1.916 \pm 0.018$ | 1.73 |
| $g_z$ | $1.984 \pm 0.007$ | 1.96 |

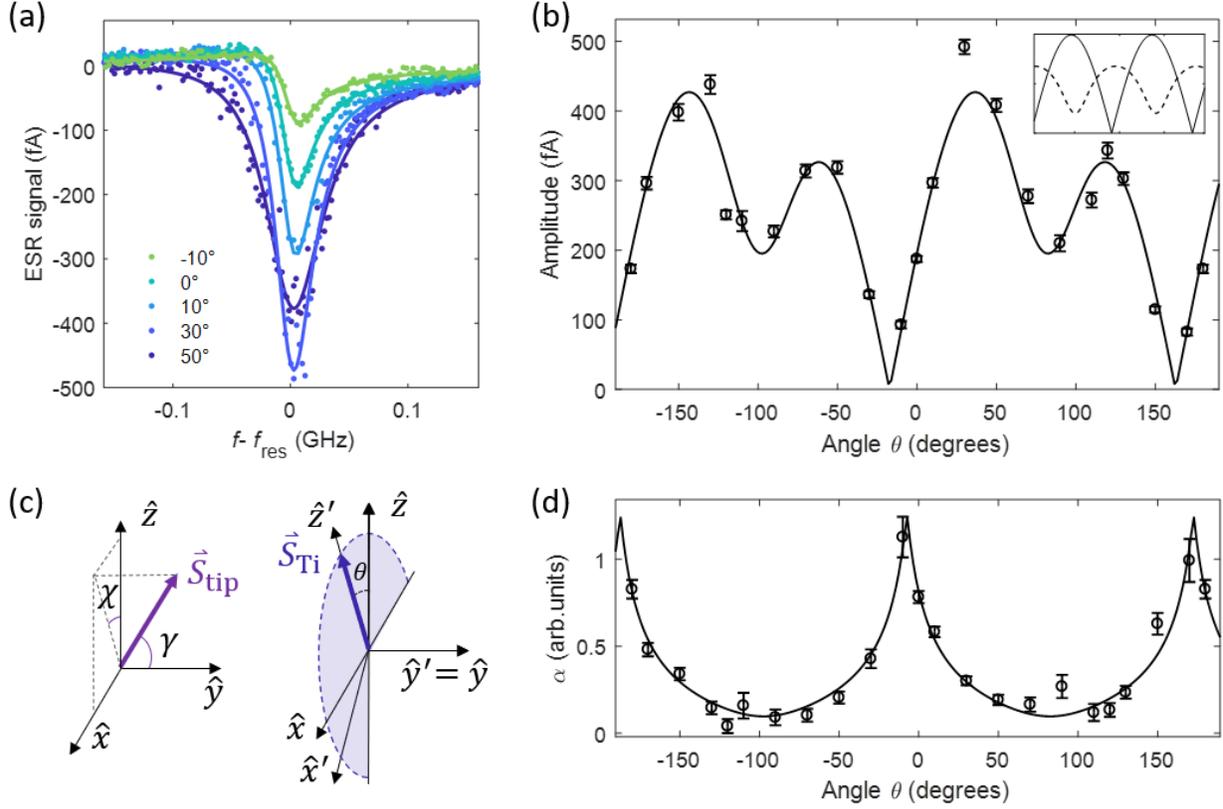

FIG. 3. ESR amplitude at different vector fields for Ti$_h$. (a) ESR spectra at different angles ($\theta$) of the external magnetic field ($V_{DC}$ = 40 mV, $I$ = 20 pA, $V_{RF}$ = 10 mV, $T$ = 1.12 K, $B_{ext}$ = 0.7 T). (b) ESR amplitude as a function of $\theta$. The ESR amplitude of each point is extracted from the fit of ESR curves to Eq. 1, where all the ESR curves are measured at $V_{DC}$ = 40 mV, $I$ = 20 pA, $V_{RF}$ = 10 mV, $B_{ext}$ = 0.7 T. The solid line is a fit to the model described in the main text and Supplementary Sec. 7. Inset: Contributions from the DC magneto-conductance (solid line) and the driving factor $\Phi(\Omega)$ (dotted line) to the fit. (c) Schematics of the relative spin orientations of tip (left) and sample (right). We define $x$- and $z$-axes as in-plane and out-of-plane directions of the sample surface, respectively, as defined in Fig. 1(a). The direction of the tip spin is denoted by the angles $\gamma$ and $\chi$ with respect to the y- and z-axes in the lab frame. At the right coordinate, $z'$-axis is the quantization axis of the Ti spin, lying in the $xz$-plane in the lab frame since we rotate $\boldsymbol{B}_{ext}$ in that plane ($\boldsymbol{y'} = \boldsymbol{y}$). (d) Asymmetry factor $\alpha$ as a function of $\theta$. The asymmetry factor $\alpha$ is extracted from the same fit in (b). It describes the relative contribution of homodyne detection over the DC magneto-conductance, depending on the alignment of the tip spin and Ti spin.

We will now focus on the amplitude dependence of ESR-STM on Ti. Figs. 1(d) and 3(a) show two examples of ESR spectra for Ti measured with different tips as a function of the magnetic field direction $\theta$, which show rather dramatic changes of the amplitude of the ESR signals as a function of $\theta$. This variation is a consequence of the change of the alignment between the Ti spin and the spin of the magnetic cluster at the tip. Importantly, the spin at the tip plays two different roles based on the piezoelectric coupling model in the ESR measurement: (1) providing the oscillating magnetic field to drive the ESR transitions [12] and (2) detecting the ESR signals based on the tunneling magneto-resistance effect [6,18].

In the piezoelectric coupling model, the Ti atoms are moving due to the applied RF electric fields and thus experience a time-varying magnetic field under the spatially inhomogeneous effective magnetic field, $\boldsymbol{B}_{\text{tip}}$, created by the magnetic interaction between the tip and Ti [12]. Here, we assume that the tip spin has a strong easy-axis magnetic anisotropy, which keeps the tip spin aligned in a certain (fixed) direction when $\boldsymbol{B}_{\text{ext}}$ changes. However, we consider that the tip spin flips by 180 degrees as needed to minimize its Zeeman energy (i.e. it is pointing up or down along its easy axis) [19]. Fig. 3(c) introduces the coordinates used in the model. The $xyz$-coordinates are determined by the surface normal ($z$) and the plane of $\boldsymbol{B}_{\text{ext}}$ ($xz$-plane) as defined in Fig. 1(a), whereas the $x'y'z'$-coordinate is defined to have the $z'$-axis along the quantization axis of the Ti spin (the direction of the external magnetic field). Note that we assume $\boldsymbol{B}_{\text{ext}} + \boldsymbol{B}_{\text{tip}} \approx \boldsymbol{B}_{\text{ext}}$ for ESR amplitude analysis.

Based on the piezoelectric coupling model, the simplified equation for the ESR amplitude $I_1$ is given by

$$I_1 \propto \Phi(\Omega) \cdot \left[\langle \vec{S}_{\text{tip}} \rangle \cdot \langle \vec{S}_{\text{Ti}} \rangle\right], \qquad (2)$$

where $\Omega$ is the Rabi rate, $\Phi(\Omega) = \Omega^2 T_1 T_2 / (1 + \Omega^2 T_1 T_2)$ is the driving factor on resonance, which ranges from 0 to 1 depending on how far the spin population is driven into equal state population [20]. Eq. (2) shows that the ESR signals are determined by the product of the driving strength ($\Phi(\Omega)$) and the spin contrast for the detection ($\langle \vec{S}_{\text{tip}} \rangle \cdot \langle \vec{S}_{\text{Ti}} \rangle$).

Both driving and detection vary when rotating the magnetic field. When we define the quantization axis of the Ti spin $\boldsymbol{S}_{\text{Ti}}$ as the $z'$-direction, the $x'$- and $y'$-components of $\boldsymbol{B}_{\text{tip}}$ provide the driving magnetic field $\boldsymbol{B}_1$ of ESR transitions, where we choose $y' = y$. Since the Ti spin is aligned along the external magnetic field, rotating this field in the *xz*-plane results in a corresponding change of the amplitudes of the $x'$- and $z'$-components of $B_{tip}$ as experienced by the Ti atom on the surface. The strength of $B_1$ consequently changes depending on the amplitude of the $x'$-component of $B_{tip}$, while the $y'$-component of $B_{tip}$ provides a constant (independent of the angle $\theta$) $B_1$ field. Since $\Omega$ is proportional to the amplitude of the driving magnetic field, the driving factor in Eq. (2) varies accordingly (see the dashed line in the inset of Fig. 3(b)).

The detection of ESR signals in STM is based on the tunneling magneto-resistance effect. In STM, we measure the time-averaged tunneling current [3], which is detected by the applied DC and RF bias voltages [21]. The ESR signal detected by $V_{DC}$ corresponds to the change in the time-averaged population of spin states detected by the component of the tip spin that is parallel to the quantization axis (the $z'$-component of $S_{tip}$). The second part of the ESR signal, which is detected by $V_{RF}$, results from the rectification of the applied $V_{RF}$ by the spin precession at the same frequency [22,23]. This oscillating conductance is detected by the $x'$ and $y'$ components of $S_{tip}$. For the ESR signal component that corresponds to the DC magneto-conductance, we have four different peaks when $\theta$ varies by 360 degrees, since the detection is maximum when $S_{tip}$ is parallel to the quantization axis of Ti atom ($S_{Ti}$) and the driving is maximum when $B_{tip}$ is perpendicular to $S_{Ti}$. For the AC magneto-conductance component (or homodyne detection), we have two maxima in the ESR amplitude since both driving and detection are maximum when $S_{tip}$ is perpendicular to $S_{Ti}$. A sum of DC detection and homodyne detection, thus, results in the four maxima observed in Fig. 3(b).

The amount of homodyne signals over the DC detection can be recognized from the asymmetry factor (Fig. 3d). The asymmetry of the ESR lineshape originates from the homodyne detection of the tunneling current [21,23]. The maximum of the asymmetry factor, thus, corresponds to the maximum of the $x'$ and $y'$ components of $S_{tip}$, which allows us to determine the direction of $S_{tip}$.

Interestingly, when we assume that $B_{tip}$ is parallel with the spin polarization of the tip $S_{tip}$, we are unable to produce different height of the peaks shown in Fig. 3(b) using Eq. (2). Instead, having $B_{tip}$ slightly different from the $S_{tip}$ direction yields the best fit as given by the solid line in Fig. 3(b) (see the detailed explanation in Supplemental sec. 8). The sharp dips in Fig. 3(b) appear at ~-18° and ~162°, where the DC magneto-conductance is minimum and the tip spin flips by 180 degrees. On the other hand, the rounded-out dips correspond to minima in the driving field. As the total ESR amplitude stems from a multiplication of the ESR driving and ESR detection (Eq. 2), the strongest ESR signals are found where we have high driving strength and large magneto-conductance at the same angle. Since we chose this particular tip for large ESR signal at the angle of $\theta = 50°$, where we prepared this tip, it shows the biggest signal near 50°, which means the tip's spin direction was selected depending on this starting angle. We found a remarkable degree of reproducibility of the basic features of this angle dependence using different tips and different Ti atoms (Supplemental sec. 8).

We employed a stereoscopic way to determine the $g$-factors of single Ti atoms on low-symmetry binding sites and found three markedly different principal components of the Zeeman energy. This is further evidence that 3d transition metal atoms on MgO behave dramatically differently from the free radicals often used in ensemble ESR where the variations of the g-factor are typically orders of magnitude. We found a rather pronounced variation of the ESR amplitude and lineshape with the angle of the magnetic field, which could be successfully modeled based on an intuitive model based on the piezoelectric effect. This enabled us to determine the tip's spin direction as well as the ESR driving fields in three-dimensional space and suggests that the piezoelectric effect underlies the ESR mechanism of Ti on MgO. We believe that this data set will enable predictions of novel ESR active spin centers in STM on different substrates as well as in other quantum-nanoscience platforms.

Supplementary Materials for

# Spin Resonance Amplitude and Frequency of a Single Atom on a Surface in a Vector Magnetic Field

Jinkyung Kim,[1,2,†] Won-jun Jang,[4,†] Thi Hong Bui,[1,2] Deung-jang Choi,[5,6,7] Christoph Wolf,[1,3] Fernando Delgado,[8] Denis Krylov,[1,3] Soonhyeong Lee,[1,3] Sangwon Yoon,[1,3] Christopher P. Lutz,[9] Andreas J. Heinrich,[1,2,*] and Yujeong Bae[1,2,*]

[1]Center for Quantum Nanoscience (QNS), Institute for Basic Science (IBS), Seoul 03760, South Korea

[2]Department of Physics, Ewha Womans University, Seoul 03760, South Korea

[3]Ewha Womans University, Seoul 03760, Republic of Korea

[4]Nano Electronics Lab, Samsung Advanced Institute of Technology, Suwon 13595, South Korea

[5]Centro de Física de Materiales CFM/MPC (CSIC-UPV/EHU), 20018 San Sebastián, Spain

[6]Donostia International Physics Center (DIPC), 20018 Donostia-San Sebastian, Spain

[7]Ikerbasque, Basque Foundation for Science, 48013 Bilbao, Spain

[8]Instituto de estudios avanzados IUDEA, Departamento de Fisica, Universidad de La Laguna, 38203, Tenerife, Spain

[9]IBM Almaden Research Center, San Jose, CA 95120, USA

[†]These authors contributed equally to this work.

*Corresponding authors: A.J.H. (heinrich.andreas@qns.science), Y.B. (bae.yujeong@qns.science)

**Table of contents**

S6: Extracting the anisotropy of g-factor for different atoms/tips

S7: Model of ESR signal in vector fields

S8: Model fit of ESR spectra

S9: ESR spectra at different magnitudes of magnetic fields

S10: Current dependence of ESR linewidth

1. **Experimental set up**

We performed electron spin resonance (ESR) measurements of single atoms on a surface using a home-built ultra-high vacuum (UHV) scanning tunneling microscope (STM) with high-frequency transmission cables. The system attains base temperature of 30 mK using dilution refrigerator system (Janis Research, JDR-250) which is operated with $^3$He-$^4$He gas mixture. Here, we measured ESR spectra at 1.12 K using only a small amount of mixture gas with Joule-Thomson effect. The system is equipped with superconducting magnets which provide a vector magnetic field: maximum 9T in the in-plane, 4T in the out-of-plane direction, and also rotatable in a plane with the maximum magnitude of 3T.

The radio-frequency (RF) voltage was applied to the tip-sample junction in combination with the DC voltage using a bias tee (SigaTek, SB15D2) outside of the vacuum chamber. We used different types of RF cables on the inside of the dewar, considering heat transfer at each stage. From room temperature to the 4K stage, we utilized stainless steel semi-rigid wire (COAX LTD, SC-119/50-SSS-SS) to minimize the heat transfer from the outside to the 4K stage. A superconducting NbTi cable (SC-160/50-NbTi-NbTi) is installed from the 4K stage to the Mixing Chamber (MC) which is the coldest area of the system. To transfer the cold temperature from MC to the STM head, we used a semi-rigid copper cable (SC-219/50-SC). Finally, between the semi-rigid cable and the tip, we used a flexible wire (Cooner wire, CW2040-3650P) to allow the tip to move freely.

## 2. Sample preparation

Iron (Fe) and titanium (Ti) atoms were deposited on a bilayer of MgO grown on Ag(100). The thickness of MgO layers was characterized by measuring the junction conductance at point-contact between the atomically sharp tip and Fe/Ti atoms. All the atoms on MgO were identified by the apparent height in the STM images as well as by the scanning tunneling spectroscopy (STS) spectra (Fig. S1).

Fig. S1(a) shows STM images of Fe and Ti atoms, where Ti is preferentially bonded with a hydrogen atom. In the main text, the ESR results were obtained from Ti atom sitting on the bridge site ($Ti_B$) between two nearest oxygen atoms. According to the two-fold symmetry of $Ti_B$, we marked $Ti_B$ atoms differently depending on the angle of nearest oxygen atoms: vertical ($Ti_v$) and horizontal ($Ti_h$).

Fig. S1(b) shows the difference of STS spectra of $Ti_B$ and $Ti_O$ measured using a normal and a spin-polarized tip. Note that we could not find any noticeable difference in STS spectra between $Ti_v$ and $Ti_h$. For the spin-polarized tip, 5-10 Fe atoms were transferred onto the apex of the tip until we observed electron spin resonance (ESR) signals of Ti atoms. For the hydrogenated Ti atoms with spin-1/2, the step near zero bias observed with the spin-polarized tip is an inelastic electron tunneling spectroscopy (IETS) spin excitation characteristic of a spin-1/2 atom [1]. The degree of spin-polarization of the tip was estimated from the zero-bias step [2].

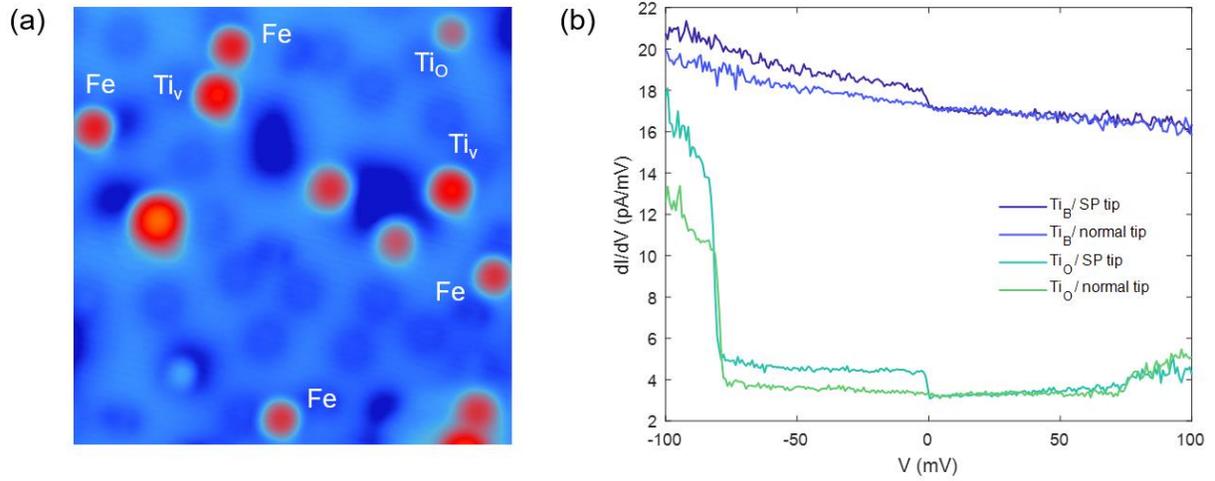

FIG. S1. (a) STM images of atoms on a bilayer of MgO (Scan size: 11.4 nm×11.4 nm, $V_{DC}$ = 100 mV, $I_{set}$ = 20 pA). $Ti_h$ and $Ti_v$ are identified using Fe and $Ti_O$ (Ti on top of oxygen) atoms as markers of oxygen atoms of MgO lattice. (b) STS spectra of $Ti_B$ and $Ti_O$ measured using the normal tip and the spin-polarized tip ($V_{DC}$ = 100 mV, $I_{set}$ = 1 nA, $B_{ext}$ = 0.5 T, $T$ = 1.12 K). Spectra for $Ti_O$ and $Ti_B$ are offset for clarity.

### 3. Point contact measurement of Ti on MgO

Fig. S2 shows the tunneling current as a function of the tip-sample distance measured by moving the STM tip closer to $Ti_B$ on MgO until the tunneling current was saturated. We define the zero distance as the position of the sudden change of slope which is attributed to contact between the tip and $Ti_B$. The point contact resistance is about 1 MΩ, showing good agreement to other references [2]. After the change of slope, the tunneling current increases more, which is presumably due to the hydrogen atom being pushed away by the STM tip.

The red curve shows the fit to the exponential function, which results in the decay constant of ~1.72 Å. We used the $I(Z)$ result to convert the current to the tip-sample distance and, thus, to fit the result in Fig. 2(a) of the main text to the model of the exchange interaction.

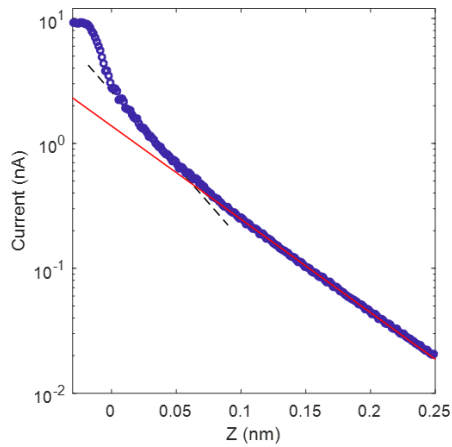

FIG. S2. Tunneling current as a function of tip-atom distance measured on $Ti_B$ on a bilayer of MgO at $V_{DC}$ = 3 mV. We define zero distance as the sudden change of the conductance slope [2].

## 4. Transfer function

In this work, we measured ESR signals with a lock-in detection of the change of the tunneling current at a certain bias voltage over a frequency sweep. While the RF signals travel from the source generator to the junction through different cables connected using several adapters and feedthroughs, we have signal losses and standing waves, which generally makes an ESR measurement in the frequency sweep mode quite difficult. To get a clear ESR signal, it is necessary to remove the frequency dependence of the RF transmission and, thus, keep the RF voltage constant at the STM junction over the entire measurement.

To cancel out the frequency-dependence of the RF voltage amplitude, the frequency-dependent transmission of the RF system is characterized, which is the transfer function [3]. Fig. S3 shows the transfer function measured at a rising step of Fe at around 14 mV in the frequency range from 5.5─30 GHz.

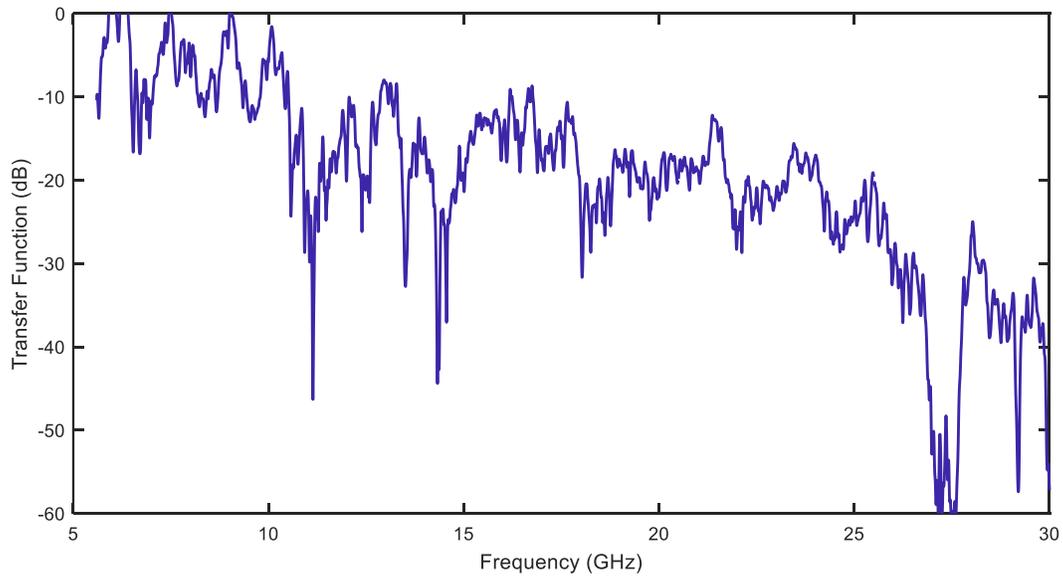

FIG. S3. RF-transfer function measured at the STM junction in the range of 5.5─30 GHz.

## 5. Density functional theory and multiplet calculations of hydrogenated Ti on MgO on Ag(100)

We used density functional theory (DFT) with pseudopotentials and plane-wave bases as implemented in Quantum Espresso (Version 6.5) [4]. Pseudopotentials to represent the ions were chosen according to the SSSP table (precision) [5], with a kinetic cutoff for the plane waves of 60 Ry and a dual of 10. The Brillouin zone integration was performed on a 3x3x1 Monkhorst-Pack k-grid and dispersive forces were treated using Grimme's van der Waals correction [6].

Following previous reports, we modeled our system as two layers of MgO frozen at the lateral lattice constant of silver ($a_{Ag}$ = 4.16 Å) expanded into a 1 nm x 1 nm lateral supercell and padded by 15 Å of vacuum in z-direction [7]. A titanium atom was added to the cell above a bridge site between two oxygen atoms and the system was relaxed until all forces were less than $10^{-3}$ (Ry/$a_0$). Then a hydrogen atom was added and the whole system was relaxed again. The relaxed structure was used to extract the coordinates and Lowdin charges used in the multiplet calculations (see TABLE SI).

We note that Ti-H does not necessarily form a perfectly perpendicular axis with the MgO surface. Slight randomization of the H position indicates that the potential energy surface for H adsorption is relatively flat which frequently results in small deviations from a perfect "on-top" absorption of H (see also the last row of TABLE SI).

| Ion | Δ electrons (#) | O-O (Å) | Mg-Mg (Å) | Z (Å) |
| --- | --- | --- | --- | --- |
| O | +0.75 | -1.36 | 0.0 | -1.27 |
| O | +0.75 | 1.38 | 0.0 | -1.27 |
| Mg | -0.73 | 0.0 | -1.67 | -2.10 |
| Mg | -0.73 | 0.0 | 1.67 | -2.10 |
| H | +0.46 | -0.03 | -0.23 | 1.76 |

TABLE SI. Point charge model used for the crystal field in the multiplet calculation. Shown are the (Lowdin) charges and cartesian coordinates for each ion in the crystal field. All coordinates are relative to the Ti adatom.

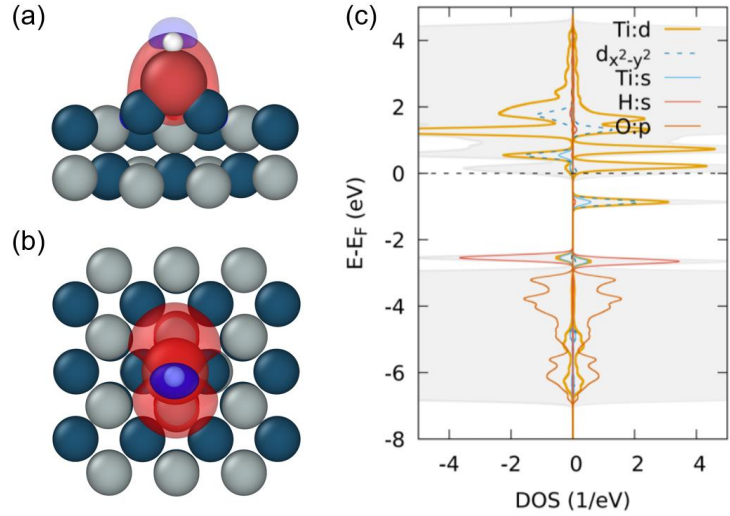

FIG. S4. (a) cross-sectional side view and (b) top view of the absorption geometry. Isosurfaces represent the spin polarization (red positive, blue negative, iso=0.0002). (c) Electronic density of states projected onto the orbitals as indicated. The oxygen 2p states are representative of the MgO valence band.

A full description of our approach to multiplet calculations for magnetic adatoms can be found in ref. [8]. In brief, we described the charge environment as point charges extracted from a DFT calculation using the Lowdin charge scheme, [9,10] and accurately treated the Ti 3d manifold using atomic wave-functions for spin-orbit coupling and electron-electron terms. The spin-orbit coupling strength was scaled to 90% of its value obtained from gas phase calculations ($\zeta$ = 18.948 meV) [11], and the 3d-3d electron repulsion was scaled to 6 eV. The overall strength of the crystal field was scaled to 10%.

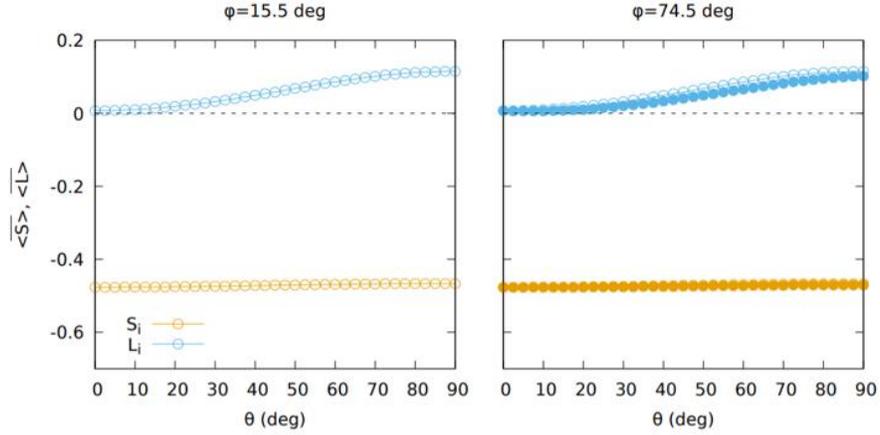

FIG. S5. Thermal averages for the expectation value of spin (<S>) and orbital moment (<L>) for two azimuthal values $\varphi = 15.5°$: $Ti_v$ and $\varphi = 75.5°$: $Ti_h$) for elevations $\theta$ measured from the in-plane direction. The anisotropy of the g-factor is the results of the anti-parallel coupling of orbital and spin moment which is more pronounced in the in-plane directions.

**Additional Calculations: hydrogenated-Ti on O-top site**

We also tested our multiplet approach for the O-top configuration. Here, the H-Ti-O bond lengths are 1.83 and 1.92 Å, respectively, with the O atom slightly displaced upward by about 0.5 Å compared to the MgO surface. Steinbrecher et al. have measured the in-plane (∥) and out of plane (⊥) g-factor of this configuration to be $g_\parallel = 1.67 \pm 0.16$ and $g_\perp = 0.61 \pm 0.09$. Using the same calculation setup described above we could reproduce these numbers to excellent agreement: $g_\parallel = 1.64$ and $g_\perp = 0.64$, see Fig. S6 and TABLE SII.

| g-factor | Reference 3 | Multiplet calculation |
|---|---|---|
| O-O | 1.67± 0.16 | 1.6356 |
| Mg-Mg | 1.67± 0.16 | 1.6356 |
| z | 0.61± 0.09 | 0.5952 |

TABLE SII. Effective g-factor for the O-top position of Ti from experiment and multiplet calculations.

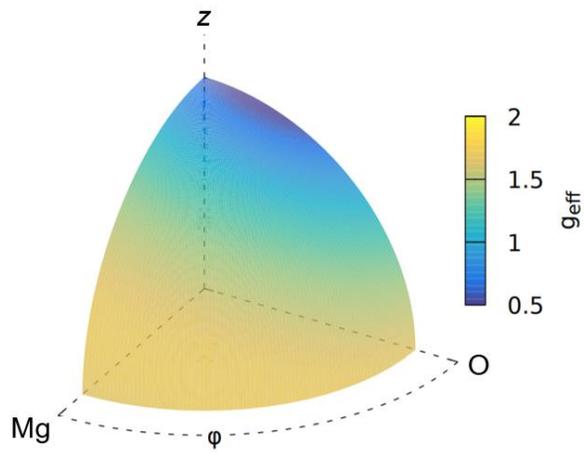

FIG. S6. Effective g-factor obtained from multiplet calculations for the O-top adsorption site. Note that in $C_{4v}$ symmetry $g_x = g_y = 1.63$.

## 6. Extracting the anisotropy of g-factor for different atoms/tips

As discussed in the main text, we observed the different magnetic moments for $Ti_v$ and $Ti_h$ at the in-plane ($\theta = 90°$) magnetic fields, while the difference at the out-of-plane ($\theta = 0°$) magnetic fields is negligibly small. At $\theta = 90°$, the magnetic moments extracted from Fig. 2(b) are $\mu_{v\|} = (0.840 \pm 0.001)\mu_B$ and $\mu_{h\|} = (0.950 \pm 0.010)\mu_B$ for $Ti_v$ and $Ti_h$, respectively. When the magnetic field is applied at $\theta = 0°$, the magnetic moments for two sites are $\mu_{v\perp} = (0.987 \pm 0.006)\mu_B$ and $\mu_{h\perp} = (0.996 \pm 0.010)\mu_B$. For spin-1/2 atomic spin, the corresponding g-factors are $g_{v\|} = 1.679 \pm 0.003$ and $g_{h\|} = 1.901 \pm 0.020$ at the in-plane field direction and $g_{v\perp} = 1.987 \pm 0.011$ and $g_{h\perp} = 1.991 \pm 0.019$ at the out-of-plane field direction.

Given that the surface normal of our sample is rotated by ~8° about y-axis and the O-O bonding direction of the MgO lattice is rotated by ~15.5° about z-axis (Fig. 1(b) in the main text), the principal axes with respect to the lab frame are given by:

$$\begin{pmatrix} x' \\ y' \\ z' \end{pmatrix} = \begin{pmatrix} \cos\beta & \sin\beta & 0 \\ -\sin\beta & \cos\beta & 0 \\ 0 & 0 & 1 \end{pmatrix} \begin{pmatrix} \cos\alpha & 0 & -\sin\alpha \\ 0 & 1 & 0 \\ \sin\alpha & 0 & \cos\alpha \end{pmatrix} \begin{pmatrix} x \\ y \\ z \end{pmatrix} \quad \text{(S1)}$$

where the rotation angles 8° and 15.5° are designated as $\alpha$ and $\beta$, respectively. Considering the rotations yields the extracted g-factors in a relationship with the g-factors along the principal axes $x'y'z'$:

$$g_{v\|} = \sqrt{(\cos\beta \cdot \cos\alpha \cdot g_{x'})^2 + (\sin\beta \cdot g_{y'})^2 + (\cos\beta \cdot \sin\alpha \cdot g_{z'})^2}$$

$$g_{h\|} = \sqrt{(\sin\beta \cdot \cos\alpha \cdot g_{x'})^2 + (\cos\beta \cdot g_{y'})^2 + (\sin\beta \cdot \sin\alpha \cdot g_{z'})^2} \quad \text{(S2)}$$

$$g_\perp = \sqrt{(\sin\alpha \cdot g_{x'})^2 + (\cos\alpha \cdot g_{z'})^2}$$

Based on Eq. S2, the calculated g-factors are $g_{x'} = 1.916 \pm 0.018$, $g_{y'} = 1.660 \pm 0.002$ and $g_{z'} = 1.984 \pm 0.007$. In the main text, the g-factors along the principal axes are given as $(g_{O-O}, g_{Mg-Mg}, g_z)$ rather than $(g_{x'}, g_{y'}, g_{z'})$ for clarity.

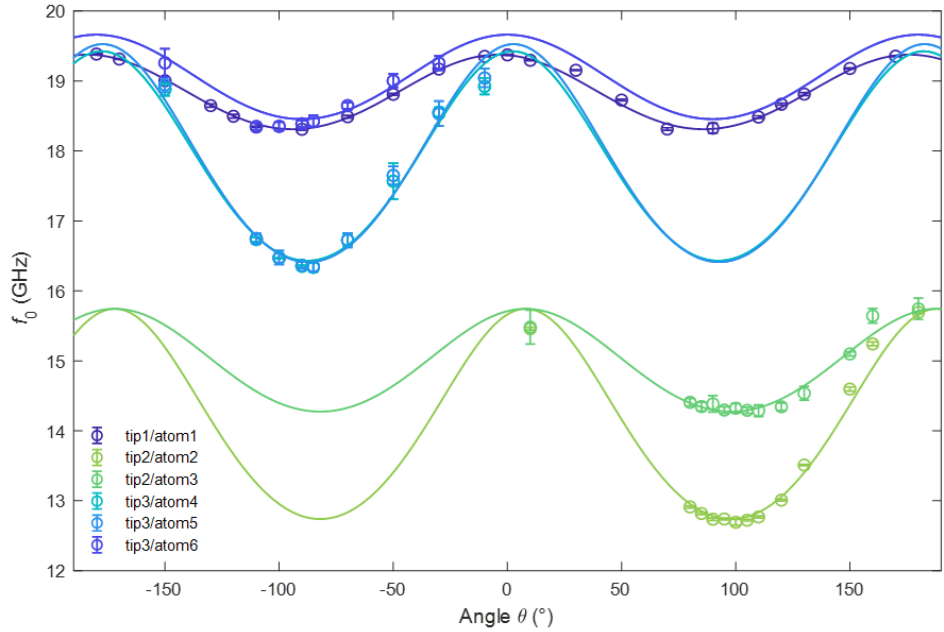

FIG.S7. $f_0$ as a function of the angle $\theta$. ESR frequencies as a function of the angle $\theta$ were measured for six $Ti_B$ atoms with three different tips. Atoms 1, 3, 6 are $Ti_h$ and atoms 2, 4, 5 are $Ti_v$. We measured ESR frequencies using tip 1 and tip 3 at the external magnetic field of 0.7 T and using tip 2 at 0.55 T ($V_{DC}$ = 40 mV, $V_{RF}$ = 10 mV, $I_{set}$ = 20 pA, and $T$ = 1.12 K). The solid lines correspond to the fit to the ellipse equation. The maximum values of $f_0$ are given at around 0~8° which indicates the surface normal of our sample is not perfectly aligned along the out-of-plane magnetic field direction ($\theta$ = 0°). Here, the small deviation of the offset angles can be attributed to different local environments such as nearby defects of MgO.

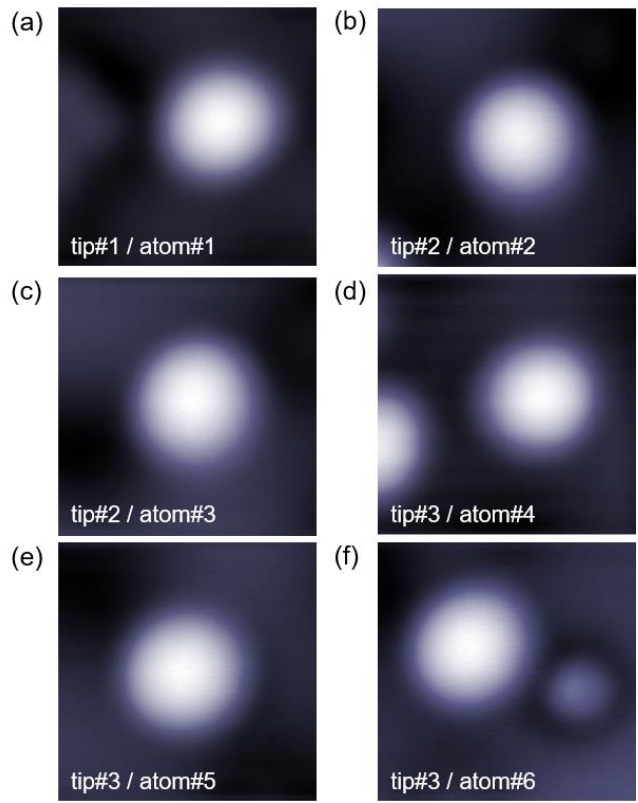

FIG. S8. STM topographic images of each atom measured at 20 pA (a, b, e) or 10 pA (c, d, f) ($V_{DC}$ = 40 mV, $V_{RF}$ = 10 mV, and $T$ = 1.12 K). Nearby atoms (d) and defects (f) can cause a subtle shift of the resonance frequency.

### 7. Model of ESR signal in vector fields

In this section, we derive the mathematical expression of ESR signals as a function of the angle of the external magnetic field. The ESR-STM spectra were obtained by measuring the tunnel current between a spin-polarized tip and individual Ti spins on the surface. The tunneling conductance was determined by the relative alignment of the tip's spin polarization direction and the Ti spin on the surface due to the magneto-resistance effect at the tunnel junction [12]

$$G = G_\text{j}(1 + a\langle \mathbf{S}_\text{tip} \rangle \cdot \langle \mathbf{S}_\text{Ti} \rangle), \tag{S3}$$

where $a$ is a normalization factor, $G_\text{j}$ is the spin-averaged junction conductance, and $\langle \mathbf{S}_\text{tip} \rangle$ and $\langle \mathbf{S}_\text{Ti} \rangle$ are expectation values of tip and Ti spin in the same reference frame (taken as the rotating frame in the following). In the rotating frame, the expectation value of the Ti spin ($\langle \mathbf{S}_\text{Ti} \rangle$) is stationary and given by the steady state solution of the Bloch equation [11]:

$$\langle S_\text{Ti}^{x'} \rangle = \langle S_\text{Ti}^0 \rangle \frac{-2\pi(f - f_\text{res})\Omega T_2^2}{1 + 4\pi^2(f - f_\text{res})^2 T_2^2 + \Omega^2 T_1 T_2}$$

$$\langle S_\text{Ti}^{y'} \rangle = \langle S_\text{Ti}^0 \rangle \frac{\Omega T_2}{1 + 4\pi^2(f - f_\text{res})^2 T_2^2 + \Omega^2 T_1 T_2} \tag{S4}$$

$$\langle S_\text{Ti}^{z'} \rangle = \langle S_\text{Ti}^0 \rangle \frac{1 + 4\pi^2(f - f_\text{res})^2 T_2^2}{1 + 4\pi^2(f - f_\text{res})^2 T_2^2 + \Omega^2 T_1 T_2}$$

where the quantization axis $z'$ of the rotating frame is defined to be the direction of the static external magnetic field $B_\text{ext}$, the $x'$-axis follows the driving RF magnetic field $B_1$ (this will be important in the following), $\langle S_\text{Ti}^0 \rangle$ is the expectation value of the Ti spin in the absence of RF voltage, $T_1$ is the spin relaxation time, $T_2$ is the spin coherence time, $\Omega$ is the Rabi rate (in angular frequency), $f$ is the frequency of the RF voltage, and $f_\text{res}$ is the resonance frequency. The tip spin $\langle \mathbf{S}_\text{tip} \rangle$, while stationary in the lab frame, rotates in the $x'y'$-plane in the rotating frame at frequency $f$. Written out explicitly, the tip spin $\langle \mathbf{S}_\text{tip} \rangle$ in the rotating frame is a combination of a constant $z'$-component $\langle S_\text{tip}^{z'} \rangle$ and a rotating $x'y'$-component:

$$\langle \mathbf{S}_\text{tip} \rangle = \langle S_\text{t}^{z'} \rangle \hat{z}' + \langle S_\text{t}^{x'y'} \rangle [\cos(-2\pi f t + \psi)\hat{x}' + \sin(-2\pi f t + \psi)\hat{y}'], \tag{S5}$$

as shown in Fig. S9. Here, the amplitude of the tip spin component in the $x'y'$-plane is denoted by $\langle S_{\text{tip}}^{x'y'} \rangle$ and its initial phase with respect to the $x'$-axis is denoted by $\psi$. Eqs. S3-S5 determine the tunneling magneto-conductance for our system.

To drive and to detect ESR signals, a RF bias voltage is applied to the STM tunnel junction in addition to the conventional DC bias voltage. The total applied bias voltage is

$$V = V_{\text{DC}} + V_{\text{RF}} \cos(2\pi f t + \varphi), \tag{S6}$$

where $\varphi$ is a phase that is not determined at the outset. Combining the above formulas, the tunnel current reads

$$I(t) = GV = G_j\big(1 + a\langle \mathbf{S}_{\text{tip}} \rangle \cdot \langle \mathbf{S}_{\text{Ti}} \rangle\big)[V_{\text{DC}} + V_{\text{RF}} \cos(2\pi f t + \varphi)], \tag{S7}$$

The high-gain low-bandwidth preamplifier of the tunnel current filters out its RF components, and the resultant averaged tunnel current is

$$\overline{I(t)} = G_j V_{\text{DC}}\left(1 + a\langle S_{\text{Ti}}^{z'} \rangle \langle S_{\text{tip}}^{z'} \rangle\right) + \tfrac{1}{2} G_j V_{\text{RF}} a \langle S_{\text{tip}}^{x'y'} \rangle \left[\cos(\psi + \varphi)\langle S_{\text{Ti}}^{x'} \rangle + \sin(\psi + \varphi) \langle S_{\text{Ti}}^{y'} \rangle\right]. \tag{S8}$$

Using a lock-in amplifier, the ESR signal $\Delta I$ is obtained by measuring the difference of the averaged tunnel current $\overline{I(t)}$ between RF on ($V_{\text{RF}}, \Omega \neq 0$) and off ($V_{\text{RF}} = \Omega = 0$, and so $\langle \mathbf{S}_{\text{Ti}} \rangle = \langle S_{\text{Ti}}^0 \rangle \hat{z}'$) [3]:

$$\Delta I = G_j a \left\{ V_{\text{DC}} \langle S_{\text{tip}}^{z'} \rangle (\langle S_{\text{Ti}}^{z'} \rangle - \langle S_{\text{Ti}}^0 \rangle) + \frac{V_{\text{RF}}}{2} \langle S_{\text{tip}}^{x'y'} \rangle \left[\cos(\psi + \varphi)\langle S_{\text{Ti}}^{x'} \rangle + \sin(\psi + \varphi) \langle S_{\text{Ti}}^{y'} \rangle\right] \right\}. \tag{S9}$$

The measured ESR signal thus has two contributions: the first (DC) contribution (first term in the equation above) originates from the change of $\langle S_{\text{Ti}}^{z'} \rangle$ under the RF drive and is sensed by the static tip magnetization along the $z'$ direction ($\langle S_{\text{tip}}^{z'} \rangle$), while the second (homodyne) contribution originates from nonzero $\langle S_{\text{Ti}}^{x'} \rangle$ and $\langle S_{\text{Ti}}^{y'} \rangle$ components and is sensed by the rotating tip magnetization (in this rotating frame description) in the $x'y'$-plane.

The expression for the Rabi rate $\Omega$ and the phase relation between $\psi$ and $\varphi$ can be obtained by considering the ESR driving process. In our model, the driving mechanism for ESR-STM on Ti/MgO is assumed to originate from electric-field-induced piezoelectric oscillations of the Ti atom on MgO in a spatially-varying tip magnetic field [13-15]. Under the applied RF voltage $V_{\text{RF}} \cos(2\pi f t + \varphi)$, oscillations of the tip-atom separation can be generally expressed as

$$z(t) = z_0 + \kappa V_{RF} \cos(2\pi f t + \varphi + \beta), \tag{S10}$$

where $\kappa$ is a piezoelectric coefficient and $\beta$ characterizes a possible phase delay between the RF drive voltage and the piezoelectric motion. The tip magnetic field originates from an exchange coupling between the tip and Ti spins:

$$H = J(z) \langle \boldsymbol{S}_{tip} \rangle \cdot \langle \boldsymbol{S}_{Ti} \rangle = \boldsymbol{B}_{tip} \cdot \langle \boldsymbol{S}_{Ti} \rangle, \tag{S11}$$

where the tip magnetic field is $\boldsymbol{B}_{tip} = J(z) \langle \boldsymbol{S}_{tip} \rangle$. In the presence of the spatially-varying magnetic field $\boldsymbol{B}_{tip}$, an oscillating Ti atom experiences a time-varying RF magnetic field

$$\boldsymbol{B}_{tip}(z(t)) = \boldsymbol{B}_{tip}(z_0) + \frac{\partial \boldsymbol{B}_{tip}}{\partial z} \kappa V_{RF} \cos(2\pi f t + \varphi + \beta), \tag{S12}$$

Here, the oscillating component perpendicular to the $\boldsymbol{B}_0$ direction (that is, in the $x'y'$ plane of the rotating frame) provides the RF drive magnetic field for the ESR transition:

$$\boldsymbol{B}_1 = \kappa V_{RF} \cos(2\pi f t + \varphi + \beta) \frac{\partial B_{tip}^{x'y'}}{\partial z} \kappa V_{RF} \cos(2\pi f t + \varphi + \beta), \tag{S13}$$

Inserting Eqs. S11 and S5 into Eq. S13 yields the expression for $\boldsymbol{B}_1$ in the rotating frame:

$$\begin{aligned}
\boldsymbol{B}_1 &= \kappa V_{RF} \cos(2\pi f t + \varphi + \beta) \frac{\partial J(z)}{\partial z} \langle S_{tip}^{x'y'} \rangle \\
&= \kappa V_{RF} \cos(2\pi f t + \varphi + \beta) \frac{\partial J(z)}{\partial z} \langle S_{tip}^{x'y'} \rangle [\cos(-2\pi f t + \psi) \hat{x}' + \sin(-2\pi f t + \psi) \hat{y}'] \\
&\approx \frac{\kappa V_{RF}}{2} \frac{\partial B_{tip}}{\partial z} \frac{\langle S_{tip}^{x'y'} \rangle}{\langle S_{tip} \rangle} [\cos(\varphi + \beta + \psi) \hat{x}' + \sin(\varphi + \beta + \psi) \hat{y}']
\end{aligned} \tag{S14}$$

In the last equation, we used trigonometrical identities and ignored the fast oscillation terms at $2f$ (the rotating wave approximation). Because the $x'$ axis of the rotating frame is defined to follow the $\boldsymbol{B}_1$ direction, the following phase requirements need to hold:

$$\psi = -\varphi - \beta, \tag{S15}$$

Physically, this means that the $x'$ axis of the rotating frame needs to be chosen in such a direction that the tip spin angle $\psi$ counters the additional phases $\varphi + \beta$ introduced during the RF magnetic field generation process. This ensures that $\boldsymbol{B}_1$ follows the $x'$ axis. Using this phase relation, we can express the RF driving field and the Rabi rate as

$$\boldsymbol{B}_1 = \frac{\kappa V_{\rm RF}}{2}\frac{\partial B_{\rm tip}}{\partial z}\frac{\langle S_{\rm tip}^{x'y'}\rangle}{\langle S_{\rm tip}\rangle}\hat{x}' = \Delta z \frac{\partial B_{\rm tip}}{\partial z}\frac{\langle S_{\rm tip}^{x'y'}\rangle}{\langle S_{\rm tip}\rangle}\hat{x}',$$

$$\Omega = \frac{g\mu_B}{2\hbar}B_1 = \Omega_0 \frac{\langle S_{\rm tip}^{x'y'}\rangle}{\langle S_{\rm tip}\rangle},$$

(S16)

where we defined a characteristic oscillation amplitude $\Delta z = \frac{\kappa V_{\rm RF}}{2}$ and a frame-independent Rabi rate $\Omega_0 = \frac{g\mu_B}{2\hbar}\Delta z \frac{\partial B_{\rm tip}}{\partial z}$.

Inserting Eqs. S4 and S15 into Eq. S9, the ESR signal $\Delta I$ can be written as

$$\Delta I = \frac{-G_j a \langle S_{\rm Ti}^0\rangle \Omega^2 T_1 T_2}{1 + 4\pi^2(f-f_{\rm res})^2 T_2^2 + \Omega^2 T_1 T_2}\left\{V_{\rm DC}\langle S_{\rm tip}^{z'}\rangle + \frac{V_{\rm RF}}{2\Omega T_1}\langle S_{\rm tip}^{x'y'}\rangle[2\pi(f-f_{\rm res})T_2 \cos\beta + \sin\beta]\right\}$$ (S17)

To cast $\Delta I$ into a more convenient form for fitting, we introduce the normalized frequency $\delta$ and its full width at half-maximum (FWHM) $\Gamma$ as

$$\delta = \frac{f-f_{\rm res}}{\Gamma/2}, \qquad \Gamma = \frac{1}{\pi T_2}\sqrt{1+\Omega^2 T_1 T_2},$$ (S18)

and the ESR signal can be rewritten as

$$\Delta I = -\frac{G_j a V_{\rm DC}\langle S_{\rm Ti}^0\rangle\langle S_{\rm tip}^{z'}\rangle \Omega^2 T_1 T_2}{1+\Omega^2 T_1 T_2}\frac{1}{1+\delta^2}\left\{1 + \frac{1}{2\Omega T_1}\frac{V_{\rm RF}}{V_{\rm DC}}\frac{\langle S_{\rm tip}^{x'y'}\rangle}{\langle S_{\rm tip}^{z'}\rangle}\left[\sin\beta + \delta \cdot \sqrt{1+\Omega^2 T_1 T_2}\cos\beta\right]\right\}$$

$$= I_1 \cdot \frac{1+\alpha\delta}{1+\delta^2}$$

(S19)

In the last equation we introduced the ESR amplitude $I_1$ and the asymmetry factor $\alpha$:

$$I_1 = -G_j a V_{\rm DC}\langle S_{\rm Ti}^0\rangle\langle S_{\rm tip}^{z'}\rangle \cdot \frac{\Omega^2 T_1 T_2}{1+\Omega^2 T_1 T_2}\cdot\left[1 + \frac{1}{2\Omega T_1}\frac{V_{\rm RF}}{V_{\rm DC}}\frac{\langle S_{\rm tip}^{x'y'}\rangle}{\langle S_{\rm tip}^{z'}\rangle}\sin\beta\right]$$

$$\alpha = \frac{\frac{1}{2\Omega T_1}\frac{V_{\rm RF}}{V_{\rm DC}}\frac{\langle S_{\rm tip}^{x'y'}\rangle}{\langle S_{\rm tip}^{z'}\rangle}\sqrt{1+\Omega^2 T_1 T_2}\cos\beta}{1 + \frac{1}{2\Omega T_1}\frac{V_{\rm RF}}{V_{\rm DC}}\frac{\langle S_{\rm tip}^{x'y'}\rangle}{\langle S_{\rm tip}^{z'}\rangle}\sin\beta}$$

(S20)

Inserting the formula of Rabi rate $\Omega = \Omega_0 \frac{\langle S_{\rm tip}^{x'y'}\rangle}{\langle S_{\rm tip}\rangle}$ (Eq. S16) yields

$$I_1 = -G_\text{j} a V_\text{DC} \langle S_\text{Ti}^0 \rangle \langle S_\text{tip}^{z'} \rangle \frac{\Omega_0^2 T_1 T_2 \left( \frac{\langle S_\text{t}^{x'y'} \rangle}{\langle S_\text{tip} \rangle} \right)^2}{1 + \Omega_0^2 T_1 T_2 \left( \frac{\langle S_\text{t}^{x'y'} \rangle}{\langle S_\text{tip} \rangle} \right)^2} \cdot \left[ 1 + \frac{1}{2\Omega_0 T_1} \frac{V_\text{RF}}{V_\text{DC}} \frac{\langle S_\text{tip} \rangle}{\langle S_\text{tip}^{z'} \rangle} \sin\beta \right],$$

$$\alpha = \frac{\frac{1}{2\Omega_0 T_1} \frac{V_\text{RF}}{V_\text{DC}} \frac{\langle S_\text{tip} \rangle}{\langle S_\text{tip}^{z'} \rangle} \sqrt{1 + \Omega_0^2 T_1 T_2 \left( \frac{\langle S_\text{tip}^{x'y'} \rangle}{\langle S_\text{tip} \rangle} \right)^2} \cos\beta}{1 + \frac{1}{2\Omega_0 T_1} \frac{V_\text{RF}}{V_\text{DC}} \frac{\langle S_\text{tip} \rangle}{\langle S_\text{tip}^{z'} \rangle} \sin\beta}.$$

(S21)

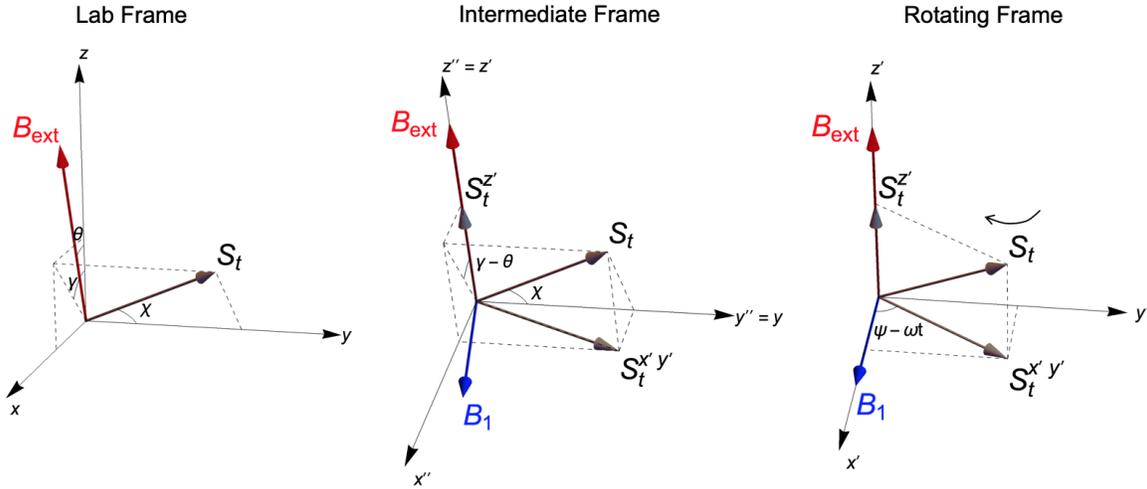

FIG. S9. Reference frames for the STM-ESR measurements.

As shown in Fig. S9, during experiments we rotate the external magnetic field $B_\text{ext}$ by an angle $\theta$ in the $xz$-plane of the lab frame while keeping the magnetic field magnitude fixed. To fit the $\theta$-dependent ESR spectra, the remaining job is to express Eq. S21 using the experimental tuning parameter $\theta$. Assuming $T_1$, $T_2$, and $\langle S_\text{Ti}^0 \rangle$ are independent of $\theta$, the only $\theta$-dependence of Eq. S21 appears in $\langle S_\text{tip}^{x'y'} \rangle$ and $\langle S_\text{tip}^{z'} \rangle$.

As the tip spin of Fe-decorated magnetic tips is in general believed to be fixed by a large magnetic anisotropy to an easy axis, its direction can be parameterized by two fixed angles $\chi$ and $\gamma$ in the lab frame (Fig. S9). We assume that changing the external magnetic field direction $\theta$ at most flips the tip spin by 180° along its easy axis (to minimize the Zeeman energy) but cannot rotate the tip spin to an arbitrary direction.

The only effect of this tip spin flipping is thus to change $\gamma$ by 180° until it satisfies $-90° \leq \gamma - \theta \leq 90°$, or $\cos(\gamma - \theta) \geq 0$. To obtain the tip spin components $\langle S_{\text{tip}}^{x'y'} \rangle$ and $\langle S_{\text{tip}}^{z'} \rangle$ in the rotating frame using $\chi$ and $\gamma$, it is convenient to consider an intermediate frame whose $z''$ axis coincides with the $z'$ axis of the rotating frame (along the $B_{\text{ext}}$ direction) and the $y''$ axis is fixed to the $y$ axis of the lab frame (Fig. S9). Since the intermediate and rotating frames share the same $z''$ ($z'$) axis, we can calculate $\langle S_{\text{tip}}^{x'y'} \rangle$ and $\langle S_{\text{tip}}^{z'} \rangle$ in the intermediate frame, which (after considering the tip spin flipping effect) yields

$$\langle S_{\text{tip}}^{z'} \rangle = \langle S_{\text{tip}} \rangle \sin\chi \, |\cos(\gamma - \theta)|,$$

$$\langle S_{\text{tip}}^{x'y'} \rangle = \langle S_{\text{tip}} \rangle \sqrt{(\sin\chi \sin(\gamma - \theta))^2 + (\cos\chi)^2}.$$

(Eq. S22)

Inserting Eq. S22 into Eq. S21 yields the final equation for fitting:

$$I_1 = -G_j a V_{\text{DC}} \langle S_{\text{Ti}}^0 \rangle \langle S_{\text{tip}} \rangle \cdot \sin\chi \, |\cos(\gamma - \theta)| \cdot \frac{\Omega_0^2 T_1 T_2 \cdot [(\sin\chi \sin(\gamma - \theta))^2 + (\cos\chi)^2]}{1 + \Omega_0^2 T_1 T_2 \cdot [(\sin\chi \sin(\gamma - \theta))^2 + (\cos\chi)^2]}$$

$$\cdot \left[1 + \frac{1}{2\Omega_0 T_1} \frac{V_{\text{RF}}}{V_{\text{DC}}} \frac{1}{\sin\chi \, |\cos(\gamma - \theta)|} \sin\beta \right],$$

$$\alpha = \frac{\frac{1}{2\Omega_0 T_1} \frac{V_{\text{RF}}}{V_{\text{DC}}} \frac{\sqrt{1 + \Omega_0^2 T_1 T_2 \cdot [(\sin\chi \sin(\gamma - \theta))^2 + (\cos\chi)^2]}}{\sin\chi \, |\cos(\gamma - \theta)|} \cos\beta}{1 + \frac{1}{2\Omega_0 T_1} \frac{V_{\text{RF}}}{V_{\text{DC}}} \frac{1}{\sin\chi \, |\cos(\gamma - \theta)|} \sin\beta}.$$

(Eq. S23)

## 8. Model fit of ESR spectra

(i) Amplitude fit

In this section, we describe the fit functions of Figs. 3(b) and (d) in the main text. As proposed by Seifert et al. [17], given that no phase delay between the RF drive voltage and the piezoelectric motion ($\beta = 0$), Eq. S23 can be simplified:

$$I_1 = -G_j a V_{DC} \langle S_{Ti}^0 \rangle \langle S_{tip} \rangle \cdot \sin\chi \, |\cos(\gamma - \theta)| \cdot \frac{\Omega_0^2 T_1 T_2 \cdot [(\sin\chi \sin(\gamma - \theta))^2 + (\cos\chi)^2]}{1 + \Omega_0^2 T_1 T_2 \cdot [(\sin\chi \sin(\gamma - \theta))^2 + (\cos\chi)^2]},$$

$$\alpha = \frac{1}{2\Omega_0 T_1} \frac{V_{RF}}{V_{DC}} \frac{\sqrt{1 + \Omega_0^2 T_1 T_2 \cdot [(\sin\chi \sin(\gamma - \theta))^2 + (\cos\chi)^2]}}{\sin\chi \, |\cos(\gamma - \theta)|}.$$

(Eq. S24)

Fig. S10(a) shows the ESR amplitudes measured at different angles of the external magnetic field (same as Fig. 3(b) in the main text) and the fit to Eq. S24, which gives four peaks with the same height.

To describe the peaks with different heights, we consider a subtle difference between the spin polarization of the tip and the driving field (tip magnetization), $\boldsymbol{B}_{tip} \neq J(z) \langle \boldsymbol{S}_{tip} \rangle$. While the spin polarization in the tunneling process is determined by the spin-resolved density of states at the Fermi level [12,16], the tip magnetization for the ESR driving fields results from the total occupied states. Considering such difference introduces a slight angle offset to the driving fields (Eq. S16) with respect to the spin polarization, which rewrites $\Omega$ using $\gamma'$ and $\chi'$ rather than $\gamma$ and $\chi$:

$$I_1 = -G_j a V_{DC} \langle S_{Ti}^0 \rangle \langle S_{tip} \rangle \cdot \sin\chi \, |\cos(\gamma - \theta)| \cdot \frac{\Omega_0^2 T_1 T_2 \cdot [(\sin\chi' \sin(\gamma' - \theta))^2 + (\cos\chi')^2]}{1 + \Omega_0^2 T_1 T_2 \cdot [(\sin\chi' \sin(\gamma' - \theta))^2 + (\cos\chi')^2]},$$

$$\alpha = \frac{1}{2\Omega_0 T_1} \frac{V_{RF}}{V_{DC}} \cdot \frac{\sqrt{(\sin\chi \sin(\gamma - \theta))^2 + (\cos\chi)^2} \cdot \sqrt{1 + \Omega_0^2 T_1 T_2 \cdot [(\sin\chi' \sin(\gamma' - \theta))^2 + (\cos\chi')^2]}}{\sin\chi \, |\cos(\gamma - \theta)| \sqrt{(\sin\chi' \sin(\gamma' - \theta))^2 + (\cos\chi')^2}}.$$

(Eq. S25)

Setting a small offset angle of $\boldsymbol{B}_{tip}$ and $\boldsymbol{S}_{tip}$ directions allows us to have different height of ESR amplitude peaks (Fig. S10(b)).

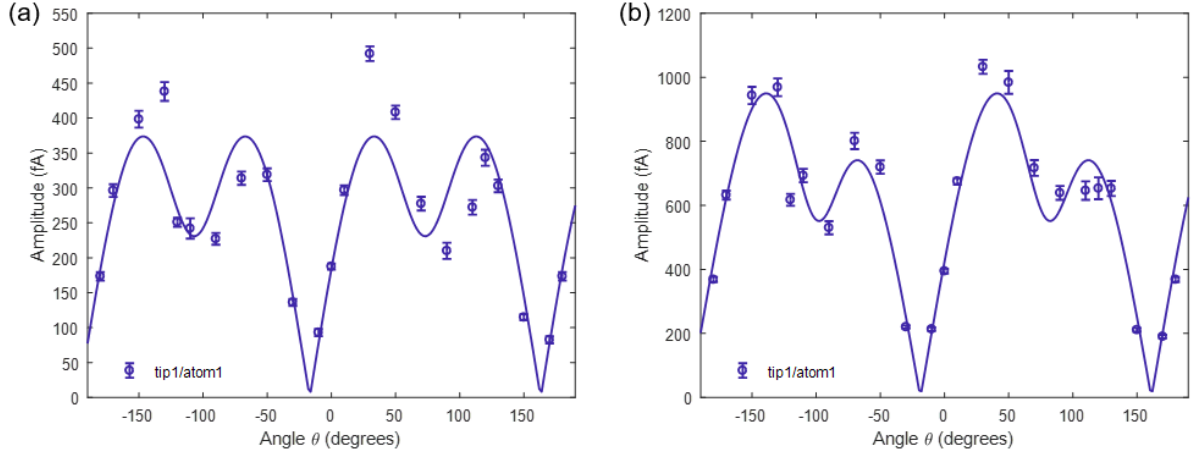

FIG.S10. (a) ESR amplitude as a function of the external magnetic field angle $\theta$ with the fit to Eq. S24, with same dataset as Fig 3(b) in the main text. (b) ESR amplitude as a function of the external magnetic field angle $\theta$ with the fit to Eq. S25, measured at 40 pA. We note that in our measurement range ($I_{set}$ = 5~40 pA), the ESR amplitude increases almost linearly with the setpoint tunneling current, maintaining the same fit shape.

| Fitting function | | | $I_1 \propto \sin\chi \, |\cos(\gamma-\theta)| \cdot \dfrac{\Omega_0^2 T_1 T_2 \cdot \left[(\sin\chi'\sin(\gamma'-\theta))^2 + (\cos\chi')^2\right]}{1+\Omega_0^2 T_1 T_2 \cdot \left[(\sin\chi'\sin(\gamma'-\theta))^2 + (\cos\chi')^2\right]}$ | | | | $I_1 \propto \sin\chi \, |\cos(\gamma-\theta)| \cdot \dfrac{\Omega_0^2 T_1 T_2 \cdot \left[(\sin\chi\sin(\gamma-\theta))^2 + (\cos\chi)^2\right]}{1+\Omega_0^2 T_1 T_2 \cdot \left[(\sin\chi\sin(\gamma-\theta))^2 + (\cos\chi)^2\right]}$ | |
|---|---|---|---|---|---|---|---|---|
| # | Type | Tip | $\gamma'$ (°) | $\gamma$ (°) | $\chi'$ (°) | $\chi$ (°) | $\gamma$ (°) | $\chi$ (°) |
| Atom 1 | $Ti_h$ | #1 | 81.59 ± 2.01 | 72.82 ± 1.32 | 69.56 ± 6.47 | 124.62 ± 4.13 | 73.11 ± 2.86 | 113.33 ± 15.41 |
| Atom 2 | $Ti_v$ | #2 | 100.44 ± 4.13 | 82.28 ± 5.61 | 97.46 ± 7.16 | 171.20 ± 2.92 | 81.65 ± 9.68 | free |
| Atom 4 | $Ti_v$ | #3 | 131.89 ± 1.83 | 132.24 ± 3.95 | 77.46 ± 3.50 | 143.53 ± 11.00 | 130.29 ± 4.70 | 77.52 ± 9.68 |
| Atom 5 | $Ti_v$ | #3 | 126.68 ± 1.60 | 134.76 ± 1.55 | 78.44 ± 2.64 | 113.73 ± 12.15 | 130.63 ± 8.02 | 76.72 ± 12.78 |
| Atom 6 | $Ti_h$ | #3 | 131.89 ± 1.83 | 132.24 ± 3.95 | 77.46 ± 3.50 | 86.12 ± 66.15 | 132.64 ± 2.69 | 76.95 ± 5.90 |

TABLE SIII: Fitting results of the angles $\gamma$, $\chi$, and $\gamma'$, $\chi'$. We used ESR data measured at 20 pA.

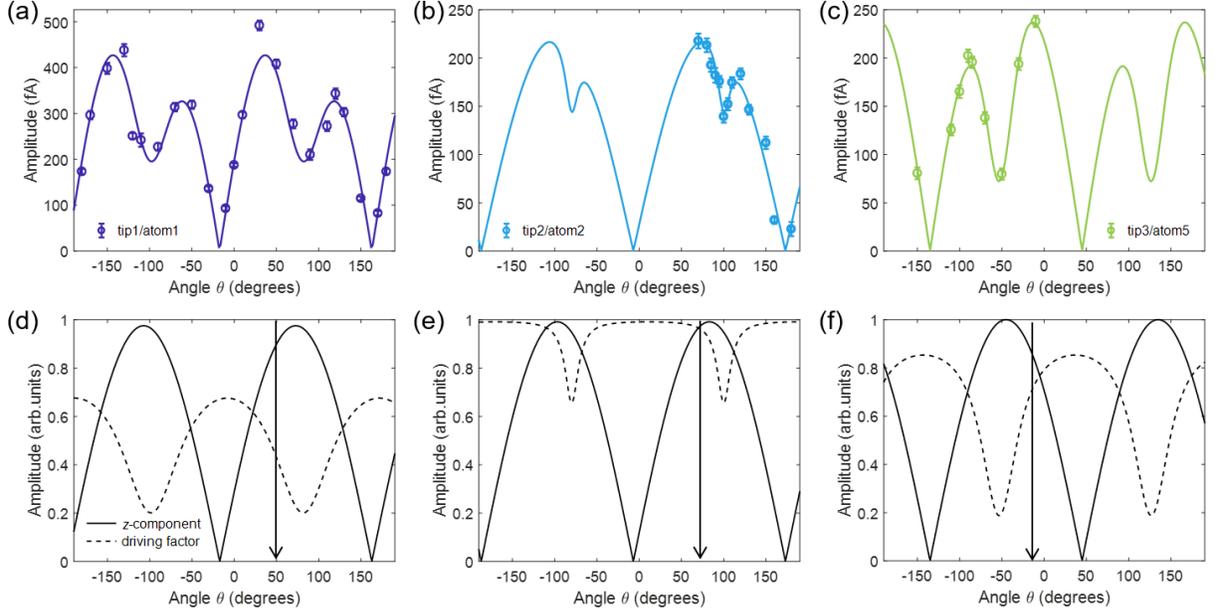

FIG. S11. (a-c) Amplitude fit and (d-f) contribution of $z'$-component of $S_{tip}$ and driving factor $\Phi(\Omega)$ to the fit (Eq. S26) for tip 1, tip 2, and tip 3, respectively. The ESR amplitude was measured on atom 1, atom 2, atom 5 as a function of angle $\theta$, respectively. Tip 2 was measured at the external magnetic field of 0.55 T and the other tips were measured at 0.7 T. The fitting results for three different tips are given in TABLE SIII. We made tip 1 at $\theta = 50°$, tip 2 at 70°, and tip 3 at -10° (arrow line). At those angles, we chose the tip maximize the ESR signal amplitude $I_1$.

In Fig. S11, the $z'$-component of $S_{tip}$ determines the overall phase of the fit and the driving factor $\Phi(\Omega) = \Omega^2 T_1 T_2/(1 + \Omega^2 T_1 T_2)$, which ranges from 0 to 1, determines the depth or sharpness of the round dips. Fig. S11 clearly shows the contribution of $S_{tip}$ ($\gamma$, $\chi$), and $B_{tip}$ ($\gamma'$, $\chi'$) in terms of driving factor $\Phi(\Omega)$, by fitting ESR amplitude as a function of the angle $\theta$.

We distinguish the direction of $S_{tip}$ ($\gamma$, $\chi$) as well as $B_{tip}$ ($\gamma'$, $\chi'$) in three-dimension using Eq. S25, where the extracted $B_{tip}$ direction is at $\gamma' = \sim 82 \pm 4°$, $\chi' = \sim 110 \pm 8°$ and the extracted spin polarity $S_{tip}$ direction at $\gamma = \sim 73 \pm 2°$ for tip 1 (see TABLE SIII). As we rotate the external magnetic fields in the $xz$-plane, the tip spin, while is at most fixed at a certain direction, flips at $\theta = \gamma \pm 90°$ to minimize the Zeeman energy, which gives the sharp dips at $\theta = \sim -18°$ and $\sim 162°$. We also confirmed that the direction of $B_{tip}$ and $S_{tip}$ varies for different tips. With tip 2, the difference between the direction of $B_{tip}$ and $S_{tip}$ is ~17°, bigger than other tips (less than ~8°). This bigger difference makes more imbalanced position of peaks in the amplitude fit (Fig. S11(e)). Based on this simple model, the amplitude of the ESR signals at different angles of $B_{ext}$ can be well described as given by the curve in Fig. S11.

From the amplitude fit presented so far, we demonstrate three points:

1. The direction of the spin polarization and the magnetization of the tip can have subtle mismatch depending on each tip. Here, we derived those directions in three dimensions from the amplitude fit.

2. The tip spin direction, which gives the largest ESR signal, is aligned in the middle of maximum $z'$- and $x'$- components of the $S_{tip}$. This ensures that ESR driving and detection are both effective.

3. The DC magneto-conductance contributes more strongly to the ESR amplitude than the AC magneto-conductance at $V_{RF} = 10$ mV.

(ii) Asymmetry factor fit

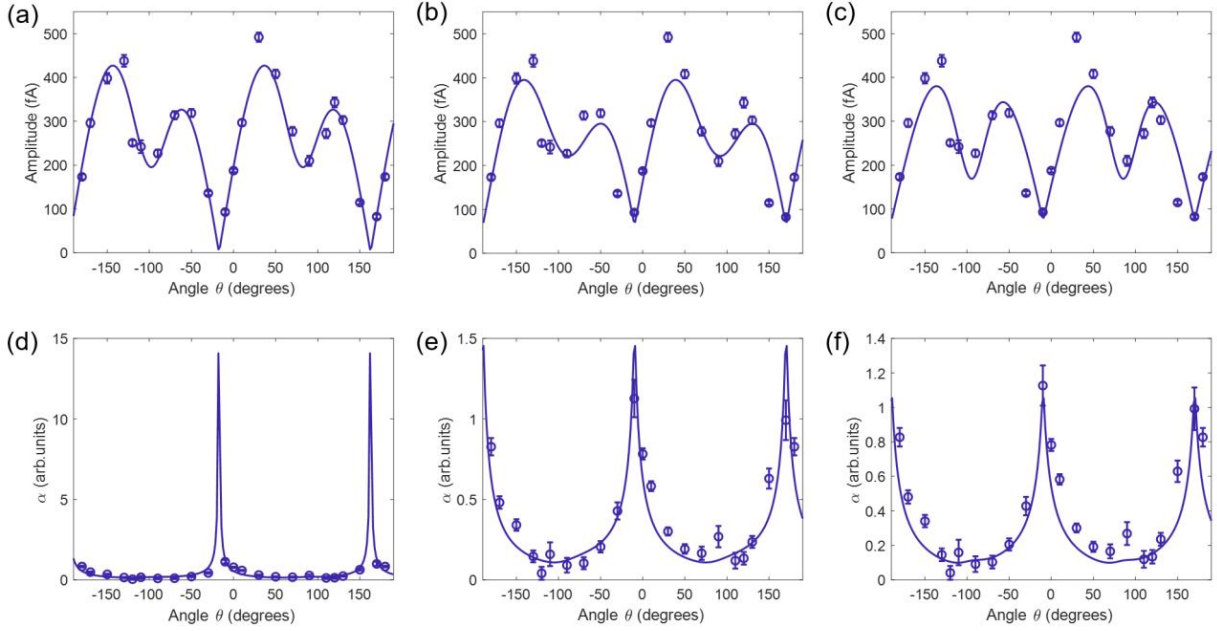

FIG.S12. The fit of ESR amplitude and asymmetry factor $\alpha$ as a function of the angle $\theta$ with tip 1. (a, b, c) and (d, e, f) shares same fitting parameters, respectively. We fit the data where $\beta$ is zero (a, d), where $\beta$ is not zero (b, c, e, f).

Using Eq. S25, we fit the ESR amplitude and the asymmetry factor α with same parameters as shown in Fig. S12. Each column corresponds to the results fitted using the same parameters. We have several different conditions to make the fit best, since the fitting equation includes many parameters. Fig. S12 shows that when $\beta = 0$, the asymmetry factor has peaks which sharply rise (Fig. S12 (d)) where $\beta$ relates how the piezo-electric motion is converted to the driving magnetic fields. It is also possible to fit our data in case of $\beta \neq 0$ (Fig. S12 (b, c, e, f)). It is challenging to make a reasonable fit for both when $\beta \neq 0$, due to the more complicated form of the fitting equation (Eq. S23). However, we note that the overall parameters are similar, which presents similar shape of the fit.

When the external magnetic field is mostly aligned with the tip spin in the $xz$-plane ($\theta = \gamma, \gamma \pm 180°$), the asymmetry factor $\alpha$ is minimum, where the DC magneto-conductance is maximum and the AC magneto-conductance and the driving field are minimum, resulting in a symmetric Lorenztian lineshape of the ESR spectrum [2,17]. On the other hand, when we rotate the external magnetic field to the perpendicular direction with the tip spin in the *xz*-plane ($\theta = \gamma \pm 90°$), the asymmetry factor α rapidly rises. At $\theta = \gamma \pm 90$, there is only homodyne detection of the Ti spin. The asymmetry factor effectively shows the trend of the relative magnitude of DC and AC detection by assuming the strong magnetic anisotropy of the tip's spin. The ESR signals measured at $\theta = -10°$ and $-110°$ are given in Fig. S13.

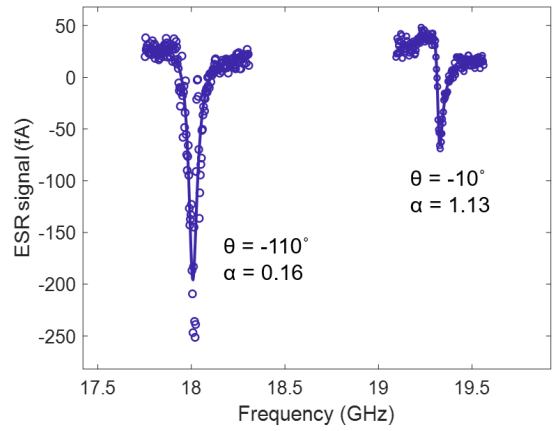

FIG. S13. Example of two ESR spectra at $\theta$ = -110° and $\theta$ = -10° using tip 1. ESR spectrum at $\theta$ = -110° (left) shows almost lorentzian shape, represented with small α. In contrast, at $\theta$ = -10° (right), ESR lineshape is more asymmetric with a larger value of *α*.

## 9. ESR spectra at different magnitudes of magnetic fields

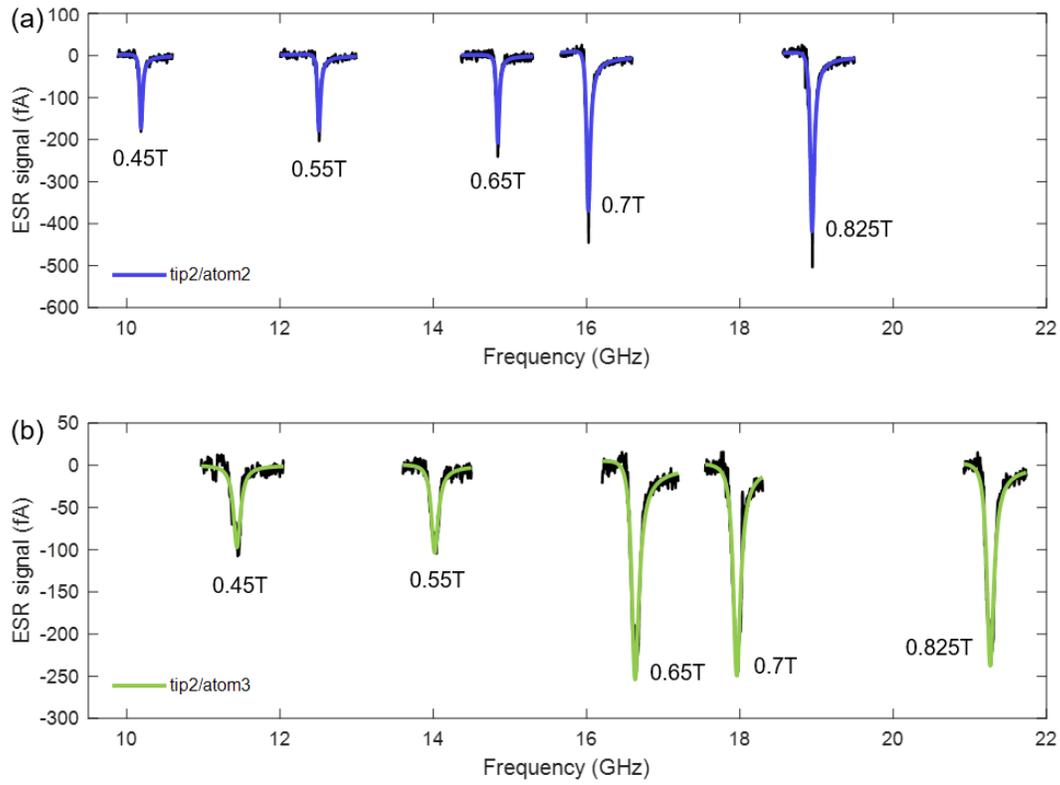

FIG.S14. ESR spectra on $Ti_v$ and $Ti_h$ with different magnitude at in-plane direction ($\theta = 90°$) with tip 2, which represents right panel at Fig. 2(b) in the main text ($V_{DC}$ = 40 mV, $V_{RF}$ = 10 mV, $I_{set}$ = 20 pA, and $T$ = 1.12 K).

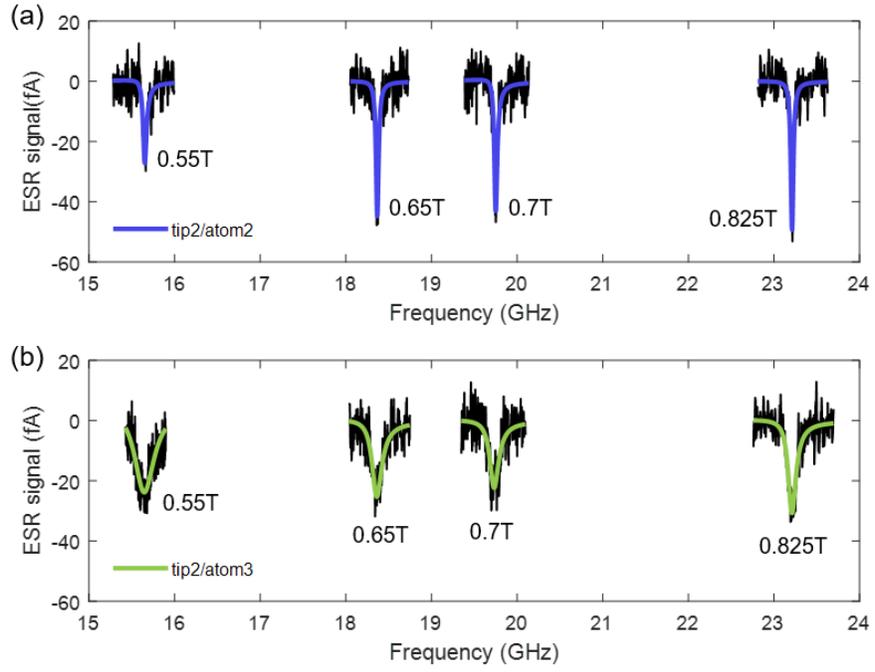

FIG.S15. ESR spectra on $Ti_v$ and $Ti_h$ with different magnitude at out-of-plane direction ($\theta = 0°$) with tip 2, which represents left panel at Fig. 2(b) in the main text ($V_{DC}$ = 40 mV, $V_{RF}$ = 10 mV, $I_{set}$ = 20 pA, and $T$ = 1.12 K).

## 10. Current dependence of ESR linewidth

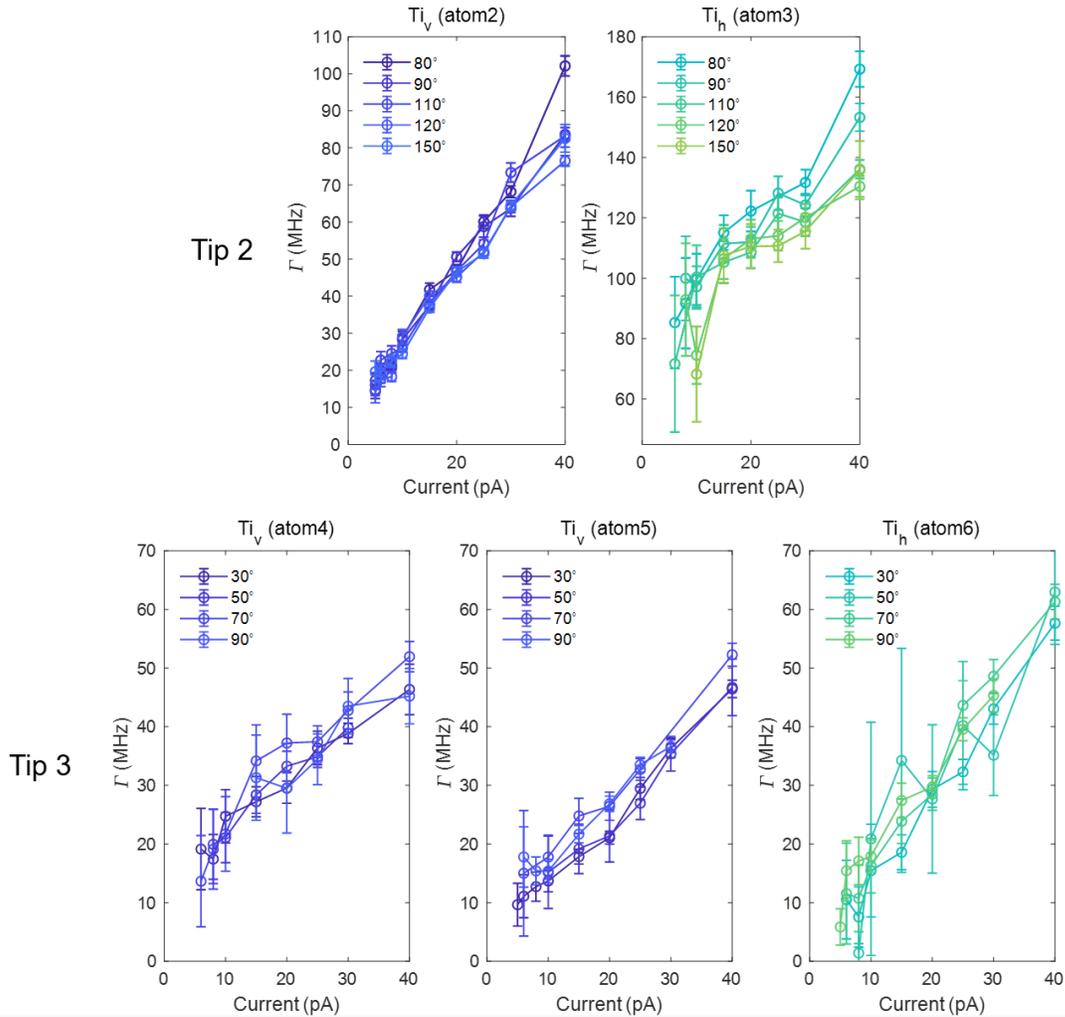

FIG.S16. Current dependence of ESR Linewidth $\Gamma$ on $Ti_v$ and $Ti_h$ with tip 2 and tip 3 at different angle $\theta$ ($V_{DC}$ = 40 mV, $V_{RF}$ = 10 mV and $T$ = 1.12 K). Color is adjusted for distinction of $Ti_v$ and $Ti_h$.

We found that linewidth $\Gamma$ of $Ti_v$ and $Ti_h$ strongly depends on the tip. When we consider the orbital contribution of $Ti_v$ and $Ti_h$ it could cause different linewidth $\Gamma$, in the interplay with the spin polarized tip.